\documentclass[draft]{agujournal2019}
\usepackage{url} 
\usepackage{lineno}
\usepackage[inline]{trackchanges} 
\usepackage{soul}
\usepackage{amssymb}
\draftfalse

\journalname{JGR: Space Physics}

\begin{document}
%
%


\title{Statistical Characteristics of the Electron Isotropy Boundary}

%
%




\authors{C. Wilkins, V. Angelopoulos, A. Runov, A. Artemyev, X-J. Zhang, J. Liu, E. Tsai}


\affiliation{1}{Department of Earth, Planetary, and Space Sciences, UCLA}




\correspondingauthor{Colin Wilkins}{colinwilkins@ucla.edu}




\begin{keypoints}
\item Using observations by the ELFIN CubeSats in 2019 and 2020 we statistically characterize the properties of 50 keV to $\sim$5 MeV electron isotropy boundaries (IBs), including occurrence rates, spatial distribution, and associated precipitating energy fluxes versus magnetic local time, latitude, and geomagnetic activity indices
\item We found that electron IBs occur over a wide range of nightside local times and latitudes under any geomagnetic conditions, and exhibit four orders of magnitude variation in electron precipitation due to geomagnetic activity. They contribute on average up to 20\% of the total high-latitude nightside $\geq$50 keV electron precipitation, at times becoming the predominant contributor to such precipitation
\item We discuss implications of IB precipitation for selected magnetospheric and ionospheric processes, global atmospheric power input, and applications of IB observations to magnetic field and particle flux modeling
\end{keypoints}

%
%

%
%


\begin{abstract}
Utilizing particle data from the ELFIN satellites, we present a statistical study of $\sim$2000 events from 2019-2020 characterizing the occurrence in magnetic local time and latitude of $\geq$50 keV electron isotropy boundaries (IBs), and the dependence of associated precipitation on geomagnetic activity. The isotropy boundary for an electron of a given energy is the magnetic latitude poleward of which persistent isotropized pitch-angle distributions ($J_{prec}/J_{perp}\sim 1$) are first observed to occur. The boundary is interpreted as resulting from magnetic field-line curvature scattering (FLCS) in the equatorial magnetosphere, a process that violates the first adiabatic invariant of particle motion. The FLCS isotropization mechanism is readily recognizable over a wide range of electron energies (10s of keV to several MeV) in the often highly dynamic transition region between the outer radiation belt and plasma sheet, where such populations are commonly found. We find that electron IBs can be well-recognized on the nightside from dusk until dawn, under all geomagnetic activity conditions, with a peak occurrence rate (averaged over all activity levels) of almost 90\% near $\sim$22 hours in magnetic local time (MLT), and remaining above 80\% from pre- to post-midnight (21 to 01 MLT). The IBs span a wide range of IGRF magnetic latitudes from $60^\circ$-$74^\circ$, with a maximum occurrence between $66^\circ$-$71^\circ$ (L of 6-8), shifting to lower latitudes and pre-midnight local times as geomagnetic activity increases. The precipitating energy flux of $\geq$50 keV electrons averaged over the IB-associated latitudes varies over four orders of magnitude, up to $\sim$1 erg/cm$^2$-s, and often includes electrons of energy exceeding 1 MeV. The local time distribution of IB-associated electron energies and precipitating fluxes also exhibit peak values near midnight for low activity, shifting toward pre-midnight for elevated activity. The percentage of the total energy deposited over the high-latitude regions ($55^\circ$ to $80^\circ$; or IGRF $L\gtrsim 3$) attributed to IBs is 10-20\%, on average, or about 10 MW of total atmospheric power input. The IB-associated electron energy deposition, while narrow in latitude, can however be up to $\sim$100\% of the total $\geq$50 keV electron energy deposition over the entire sub-auroral and auroral zone region, at times exceeding 1 GW in total atmospheric power input. Both the total IB-associated precipitating flux intensity and its relative contribution to the total precipitating energy over the sub-auroral and auroral zone increase with AE, $|$Dst$|$, and Kp. We discuss implications of these results for atmospheric energy deposition, ionospheric conductivity enhancement, magnetospheric electron losses and magnetic field mapping.

\end{abstract}

\section*{Plain Language Summary}
In Earth's magnetosphere, energetic electrons are often trapped by the geomagnetic field, which can shield the planet and orbiting satellites from potentially dangerous effects. However under certain conditions, especially on the nightside of the magnetosphere, the magnetic field can become too weak or curved to maintain such trapping, and electrons can be lost to the atmosphere, where they collide and give up their energy (i.e., precipitate). On a poleward moving satellite, the magnetic latitude at which the flux of precipitating and magnetically-reflected (locally trapped) electrons of some energy first become equal is known as its isotropy boundary (IB) latitude for that energy. This happens because at that latitude the magnetic field intensity at the magnetically conjugate equator becomes too small and thus unable to maintain the particle's cyclical motion, allowing it to scatter in its pitch-angle with respect to the field direction. The electron distribution consequently randomizes in pitch-angle and many electrons thus precipitate. This latitude varies with electron energy and local time, and with geomagnetic activity, which affects the underlying magnetic topology near Earth. Isotropy boundaries for various energies can be observed by low-altitude polar-orbiting satellite missions equipped with particle detectors able to discriminate the electron flux as a function of pitch-angle and energy, such as ELFIN. Here, we characterize statistically the properties of IBs, such as their occurrence rate versus longitude and latitude, as well as the energy they deposit into the atmosphere at different geomagnetic activity levels. This ultimately allows for better understanding and predictive capability of space weather effects, as well as for improvement in magnetic field and electron flux models.

%
%

%


%
%
%
%

\section{Introduction}
\subsection{Background}
A major objective in the study of Earth's magnetosphere-ionosphere system is to characterize the space processes leading to the precipitation of energetic charged particles. One such process is magnetic field-line curvature scattering (FLCS; occasionally referred-to as ``current-sheet scattering''), in which the increased curvature and decreased strength in the equatorial magnetic field at large distances can scatter particles above a certain energy non-adiabatically in pitch-angle, isotropizing their flux distributions. Both electrons and ions (including heavy ions) can be affected in this way, though at different distances from Earth. The result of this flux isotropization is that precipitating and mirroring fluxes become comparable (i.e., $J_{prec}/J_{perp}\sim 1$), for particle energies that have equatorial gyro-radii exceeding a critical threshold \cite{Gray1982,Sergeev1983,Birmingham1984,Delcourt1996,Martin2000,Young2002}. For electrons, such FLCS-isotropization can happen at energies as low as a few 10s of keV and upward, including the 100s of keV to multi-MeV electrons found in the vicinity of the outer radiation belts \cite{Imhof1979}. Furthermore, due to the nightside equatorial near-Earth magnetic field often exhibiting a gradual change in intensity and curvature versus radial distance near the dipole-tail transition region, it is possible for this scattering process to act on different minimum energy particles over an extended range of L-shells, magnetically mapping to an extended range of latitudes in the ionosphere \cite{Sergeev1982}. Fluxes which have been isotropized by this process can be detected by polar-orbit satellites traversing a range of field-lines connected to the equatorial scattering region. The magnetic latitude corresponding to the onset of isotropy as a function of increasing (absolute) latitude for a particular particle species and energy is known as the \emph{isotropy boundary} (or ``IB'') for that species and energy \cite{Sergeev1983}. Given its ability to rapidly fill the loss cone over a wide energy range, local time, and latitudinal range, the process of curvature scattering presents a potentially significant means by which energetic particles and their associated kinetic energy can be deposited into the atmosphere in the vicinity of IBs.\\

Contrary to proton IBs, whose properties have been well-explored previously \cite{Sergeev1982, Sergeev1983, Newell1998, Donovan2003, Yue2014, Sergeev2015b, Dubyagin2018}, the electron IBs have received less attention and their occurrence and properties have been poorly investigated \cite{Sergeev2018,Capannolo2022}. This has largely been due to observational constraints of past studies, which have contended with limited latitudinal coverage (e.g. RadSat, UARS, Van Allen Probes, GOES, Geotail, THEMIS, etc.), or issues with particle detection capability, such as insufficient electron pitch-angle resolution and energy range, dynamic range or sensitivity, uncertainties in sensor cross-calibration between look directions, and cross-species contamination (e.g. POES, DMSP). Because energetic electron IBs are found as high as $74^\circ$, exhibit orders of magnitude variation in flux and precipitation energy, and often appear poleward of proton IBs, previous studies were ill-equipped to fully characterize their properties. To address these issues, we used recently acquired data from the Electron Losses and Fields Investigation (ELFIN) satellites \cite{Angelopoulos2020}, whose observations from circular polar Low Earth Orbit (LEO) provide latitudinal coverage spanning 55$^\circ$ to 80$^\circ$, 24 hours in aggregate local time, and electron energies between $\sim$50 keV and $\sim$5 MeV, with a single, low-noise, high-sensitivity, proton-rejecting electron sensor used to measure fluxes over all pitch-angles during each spin (once per 2.8 seconds).\\

Beyond characterization of the electron IB proper, the properties and role of isotropic $\geq$100 keV electrons associated with IBs in several important magnetospheric and ionospheric processes have remained similarly veiled. For example, because FLCS-isotropized electrons can include populations of tens of keV and up to the MeV range, they can penetrate to lower altitudes into the high-latitude atmosphere than typical auroral fluxes, potentially increasing conductivity and chemical reactivity at altitudes as low as the ionospheric D-region \cite{Fang2010}. Additionally, because the FLCS isotropization mechanism can act rapidly (on the order of a bounce period) in the equatorial region where inward radial diffusion transports particles into the outer radiation belt, it can potentially prevent such particles from ever becoming trapped. These flux losses would not be accounted for as part of outer radiation belt precipitation due to being outside the trapped flux region. Such efforts are also confounded by the fact that the isotropy boundary is difficult to recognize on equatorial spacecraft traversing the tail-dipole transition region, because the loss cone at the equator is only $\sim$1$^\circ$, too small to resolve by typical particle instruments; thus their fluxes are not accounted for in energetic precipitation modeling and prediction based on such data. Further, knowledge of the IB location as a function of time can be a useful tool for near-instantaneous remote-sensing of the equatorial magnetic field configuration, and of the particle populations residing there. This can help refine equatorial magnetic field and particle flux models, especially when paired with equatorial satellites \cite{Sergeev1993, Sergeev2015a, Ilie2015, Shevchenko2010}. Here, we expand on preceding observational results, which have reported the presence of both isotropic electrons and protons with energies above typical plasma sheet auroral processes ($>$10s of keV) throughout the magnetosphere, but especially on the nightside.\\

For electrons in particular, past studies have reported isotropic precipitation structures often found within $\pm$4 hours of midnight, persisting for multiple spacecraft orbits (timescales of hours) with well-populated (filled) loss cones up to several MeV in energy \cite{Imhof1977,Imhof1979,Imhof1997}. The most striking feature of these structures was the energy-latitude dispersion in the onset of isotropy: higher energies were monotonically isotropized at lower latitudes than lower energies \cite{Imhof1997}. These combined properties suggested that the underlying generation mechanism must persist over a wide range a geomagnetic conditions, and could not be due to transient processes, such as wave-particle scattering. It was proposed that such long-lasting isotropy structures were likely the result of particle scattering by the curvature of the background magnetic field within the cross-tail current, which extends both in latitude and in local time around midnight. Similar isotropy boundary effects for protons have been reported to occur at lower latitudes, the latter characteristic explained as due to the larger gyro-radii of protons. Interestingly, intense localized peaks in energetic proton precipitation were found to often occur in close poleward proximity to the proton IB (itself close to the auroral oval), with an apparent pre-midnight peak occurrence \cite{Newell1998}. This suggests that the IB precipitation may be replenished by repeated instances of freshly accelerated and transported particles from the magnetotail. The equivalent properties for electron IBs are addressed in this work. For completeness, we note that interactions with magnetopause may give rise to similar dispersive isotropic particle scattering onto Earth-connected field lines \cite{Lyons1987}, although these cases are not the focus of this study, and have been excised from our database.\\

\subsection{Model of energetic electron isotropization}
Isotropy boundaries emerge when freshly accelerated (heated) particles convect or drift through the near-earth equatorial magnetosphere near locations of strong equatorial magnetic field curvature or weakened strength. After traversing the equator, the resulting modification of a particle's pitch-angle depends highly on the ratio of its incoming gyroradius ($r_L$) to the local scale length of variations in the background magnetic field. On the nightside in particular, as the distance from Earth increases, the most significant magnetic scale variations correspond to the smallest (sharpest) magnetic curvature radius $R_C = |\hat{\bf b}\cdot \nabla \hat{\bf b}|^{-1}$ of the field, where $\hat{\bf b} = {\bf B}/B$ is the unit tangent vector to the magnetic field ${\bf B}$. The most pronounced effects can be expected at the close Earthward vicinity of the cross-tail current, which marks a transition from a dipole-like field to an extended tail and is also accompanied by a local field strength gradient. If the particle gyroradius is much smaller than the field line curvature (i.e. $r_L \ll R_C$), the particle retains its first adiabatic invariant $\mu$ (possibly imparting an impulsive change in gyro-phase), and executes traditional guiding-center bounce and drift motions about the equatorial plane. However, if the particle gyroradius begins to approach the local equatorial radius of curvature (e.g., for higher energy particles), non-adiabatic effects emerge \cite{Gray1982,Young2002}.\\

When equatorial particle crossings become non-adiabatic due to the local geometry of the background magnetic field (as above), the resulting motion depends further on the ratio of the minimum magnetic curvature radius to the maximum equatorial particle gyroradius, defined in prevailing literature as $\kappa^2 = R_C/r_L$ \cite{Sergeev1982, Martin2000}, as well as the particle's incident gyrophase. The case of $\kappa^2 \sim 1$ results in Speiser-like motion \cite{Speiser1965}, while the range $3 \lesssim \kappa^2 \le \kappa^2_{cr}$ results in strong diffusive pitch-angle scattering, leading to isotropy upon repeated crossings of the equatorial plane (i.e. $J_{prec}/J_{perp}\sim 1$). Owing to the strong diffusion, only a few crossings are usually required to achieve isotropy. The particular critical value of $\kappa^2_{cr}$ required for efficient isotropization is independent of particle species and varies only with the magnetic field configuration in the scattering region, although at a fixed location (and field-line curvature) electrons isotropize at a higher minimum-energy than protons due to their smaller mass and smaller gyroradius. A threshold of $\kappa^2_{cr}=8$ is commonly taken as an \emph{a priori} value based on a Harris-type current sheet with constant $B_{normal}$ \cite{Gray1982,Sergeev1983}, though it can take on values between 3 and 33 over the range of possible magnetotail configurations \cite{Ilie2015}

Upon isotropization by equatorial curvature scattering, particles typically resume motion toward Earth along field lines, where they can be profiled in latitude by low-altitude polar spacecraft such as ELFIN. To provide a theoretical reference of the particle energy versus latitude at which FLCS-based isotropy could be expected to appear, we re-cast the critical $R_c \le \kappa^2_{cr} r_L$ relationship in terms of a minimum required particle kinetic energy $E_{min}^{iso}$ for isotropic field-line scattering, computing it based on model-mapped equatorial magnetic field properties at the crossing location:
\begin{equation}
E_{min}^{iso} = \left(\gamma_{min}^{iso}-1\right)mc^2 = \left[\left(1+\left(\frac{q B R_C}{\kappa^2_{cr} m c}\right)^2\right)^{1/2}-1\right]mc^2
\end{equation}
where $\gamma_{min}^{iso}$ is the minimum particle Lorentz factor corresponding to the minimum required particle energy for isotropization, $q$ is the particle's charge, $B$ is the equatorial magnetic field strength, $R_C$ is the equatorial radius of curvature, $\kappa^2_{cr}=8$ as explained above, $m$ is the mass of the particle, and $c$ is the speed of light.\\

Using this relation, an example spatial profile of minimum electron kinetic energies for efficient isotropization due to FLCS is shown in Fig. 1 over the 50 keV to 5 MeV energy range, using the combined IGRF and T89 field models \cite{Alken2021,Tsyganenko1989} at midnight local time (GSM coordinate xz-cut) for $\kappa^2_{cr} = 8$ and $K_p=2$ on 2020-09-02/14:22 UT. The profile shows that the regions capable of isotropizing 50 keV to 5 MeV electrons are confined to the equatorial plane at distances beyond several earth radii, corresponding to the dipole-tail transition from a dipole-like field to that of an extended magnetotail. The model-computed isotropy boundary field-line traces for 50 keV and 5 MeV energies are shown as blue and pink traces, respectively. \\

The above model predicts that on an equator-to-pole traversal of the pertinent field lines, as seen from a LEO polar satellite vantage point, the highest energies are isotropized first, at the lowest latitudes (smallest equatorial distances), due to their larger gyroradius and corresponding ratio to the equatorial curvature. The minimum energy of isotropization decreases monotonically as the vantage point latitude increases. This energy-latitude dispersion, characteristic of IBs, is a consequence of the field-line mapping progressively further away from Earth (especially on the nightside), where the equatorial field strength and/or radius of curvature are reduced and, as a result, the minimum required energy for isotropic scattering by Eqn. 1 is also reduced. In Fig. 1, the equatorial region corresponding to the portions of the IB dispersion ELFIN could observe is shaded yellow. The mint and light blue lines are the poleward and equatorward boundaries, respectively, of the IB dispersion region observable by ELFIN, corresponding to the ELFIN energetic particle detector's energy limits. \\

These dispersed energy-latitude isotropy signatures are the basis of IB event identification and modeling in our study. The modeled dispersion slope and minimum field-line scattering energy at a given latitude depend on the choice of magnetic field model and the quantity $\kappa^2_{cr}$. Although the data may have significant uncertainties associated with true field-line mapping, and the field model of choice may not fully represent the instantaneous magnetic topology, we assume that both IB crossings and model exhibit a monotonic energy-latitude dispersion for event-identification and modeling purposes. We note also that this model predicts that electrons exceeding the minimum required scattering energy continue to be isotropized at latitudes extending beyond their IB; however, in reality as the mapped field-line distance increases appreciably beyond the IB dispersion region, the electron dynamics can become intertwined with other plasma sheet processes.

\subsection{Outline}
In the following, we first describe the ELFIN dataset and then discuss the methods used to determine the presence of isotropy boundaries and the intensity of the associated precipitation. We next report observations of the IB occurrence rates versus MLT, L-shell, magnetic latitude and geomagnetic activity indices, alongside the observed electron energy ranges and slope of IB energy-latitude dispersion. We then discuss the computation methods and interpretation of the deposited electron energy flux associated with the IB/FLCS-dominated region. We finally summarize our findings and discuss their implications, potential applications, and the next steps in this line of investigations.

\section{Methods and dataset}
\subsection{ELFIN dataset}
We use data collected by the Energetic Particle Detector for electrons (EPD-E) instrument aboard each of the two Electron Losses and Fields Investigation (ELFIN) CubeSats. The satellites were in polar LEO (at $\sim$450 km altitude), drifting about 1 hour in magnetic local time per month. The EPD-E instruments have an energy range of approximately 50 keV to 5 MeV, sub-divided into 16 logarithmically-spaced energy channels of width less than 40\% (i.e. $dE/E \leq 0.4$), with a $22^\circ$ field of view and geometric factor of $\sim$1 cm$^2$-str. The satellites were spinning with a nominal rotation period of 2.8 s about an axis nearly perpendicular to the background magnetic field, allowing for pitch-angle determination of incident particles. The particle data collected in each spin period was subdivided in time into 16 spin sectors ($\Delta t \sim 175$ ms) by the on-board data processing unit, which were combined in ground processing with IGRF and attitude data to determine the local pitch-angle distributions. The attitude of the spacecraft typically allows for resolution of both the loss cone and locally-mirroring particle populations. Proton contamination was mitigated by an absorbant aperture foil, while side-penetrating particles were rejected by a combination of dense shielding and detector coincidence logic. The data and processing tools are publicly available using the ELFIN routines within the SPEDAS framework \cite{Angelopoulos2019}.\\

To form the statistical event database, a list of $\sim$2600 ELFIN science zone crossings ($\sim$6-7 minute data collections encompassing outer-belt and auroral zone crossings) were identified and subjected to preliminary event-selection criteria. Since the FLCS mechanism depends on the large-scale background magnetic field configuration and bulk particle transport, which can vary on the timescale of tens of minutes to hours, each science zone collection can be regarded as an instantaneous snapshot of IB properties. These science zones were collected as a part of ELFIN's Outer Belt Observations (OBO) mode, which is designed to extend between $55^\circ$ to $80^\circ$ in (absolute) magnetic latitude in a single hemisphere (L-shells $\gtrsim 3$), computed using the International Geophysical Reference Field (IGRF) model, though the start and stop limits of each individual collection can vary by up to several degrees due to the operational implementation method. The data in our study were obtained in 2019 and 2020, spanning the full range of magnetic local times (MLTs). We binned these data in 1 hour magnetic local time intervals, encompassing $>$40 collections in each MLT bin. Significantly more science zone collections were available near midnight and on the dayside than at other MLTs, requiring normalization by residence time to remove this observational bias. The collections spanned a wide range of geomagnetic activity, including storms and substorms. Because in the early part of the mission satellite spin-axis attitude was not controlled to be orbit-normal (which would provide the widest pitch-angle coverage), but, rather, science zones were targeted based on attitude being closest to ideal, further checks of our dataset were necessary to ensure the loss-cone was cleanly measured. We inspected and culled the above preliminary database to ascertain that the satellite attitude permitted sufficient pitch-angle resolution to compute precipitating-to-perp flux ratios (only a few events had to be eliminated by this process). To limit the possibility of under-counting high-latitude IB events, we also required that all science zones spanned an (IGRF-computed) L-shell exceeding $L=8$. Finally, we eliminated all events with known instrument performance issues (evident in quality flags). These constraints collectively reduced the qualifying science zones (QSZ) available for further analysis to 1922; these formed our final event database.\\

\subsection{Prototypical electron isotropy boundary crossing}
 A prototypical example of a qualifying ELFIN science zone collection containing an IB crossing is shown in Figure 2. During this $\sim$6 minute QSZ window the ELFIN-B satellite moved southward from the northern polar cap toward the magnetic equator. The top two panels show the in-situ perpendicular ($J_{perp}$) and precipitating ($J_{prec})$ energy flux spectrograms for 50 keV to 5 MeV electrons. The energy-time spectrogram of flux isotropy ratio $R_I = J_{prec}/J_{perp}$ in Panel 3, is computed from the ratio of Panels 1 and 2. Until shortly before 14:21 UT, only relatively low energy ($\leq 200$ keV) electrons are present and exhibit a high isotropy ratio ($\sim$1), suggesting a traversal of an extended electron plasma sheet. That the electrons are isotropic signifies a source location poleward of the isotropy boundary, where their energies exceed the threshold for field-line scattering. At around 14:21 UT, a rapid rise in energetic ($\geq 300$ keV) isotropized electron fluxes is observed, suggesting that the satellite entered the transition region between the inner edge of the electron plasma sheet and outer radiation belt, or tail-dipole transition region (herein dubbed the ``PS2ORB'' interface, and adhering to a specific observational definition to be introduced later). The emergence of isotropic fluxes over such a high energy range not typically found in the plasma sheet, nor associated with any known persistent wave-particle acceleration process, is consistent with the satellite being on field lines connected to a dynamical transition region dominated by field-line scattering poleward of but in close proximity to the isotropy boundary. Around 14:22 UT, an abrupt transition from isotropic to anisotropic fluxes over the energy range from 50 keV to 4 MeV is observed. Viewed in reverse-time (i.e., from increasing latitude), the latitude where flux anisotropy transitions to flux isotropy at each energy (i.e., the isotropy boundary at that fixed energy) increases as the energy decreases. This inverse (negative) latitude-energy dispersion is exactly what is expected for the isotropy boundary dispersion with energy, in which higher energies isotropize first, at lower latitudes. Based on this, we conclude that this QSZ interval contains an IB crossing.\\
 
 Panel 5 shows the isotropy boundary location (red) as a function of energy and latitude, as determined by an automated procedure. The jagged (occasionally non-monotonic) nature of the curve is a consequence of poor counting statistics at the highest energies. Also shown in the same panel is a model-based prediction of the boundary location, obtained by equatorial footpoint tracing of the ELFIN orbit with the Tsyganenko 1989 field model, and using $\kappa^2_{cr}=8$ for the isotropy boundary determination for each energy. The blue line results from a direct application of the model, which is earlier in time (hence poleward in magnetic latitude) by $\sim$15 s ($\sim$1$^\circ$) compared to observations due to model mapping uncertainties. This implies the equatorial peak cross-tail current density was apparently further away from Earth in reality at the collection time, causing the tail-dipole transition in the model to be further away from Earth and map to a higher latitude. By fitting a constant latitudinal offset to the model energy-latitude dispersion (blue line) to best match the observation (red line), we obtain the green line. The latitude-shift required to match the observations is a measure of how far the model peak cross-tail current distance from Earth deviates from the actual. For this event, the observed IB profile lasts $\sim$15 s across all energies, and is in good agreement with the modeled-then-shifted IB profile (errors less than 1 spin period of latitudinal uncertainty). The latitudinal shift required to co-locate the IB crossings provides valuable information for further investigation and model improvements that are beyond the scope of this work.\\ 
 
 After crossing the IB, the satellite moved into the outer radiation belt and subsequently in the trough (where locally-trapped relativistic electron fluxes subsided). Around 14:22:30 UT and 14:22:50 UT, near-isotropic fluxes were again observed but only at the lowest energies, $<$300 keV, despite the fact that locally-mirrored fluxes were abundant at all energies up to $\sim$3 MeV. This fact, and the lack of an energy-latitude dispersion, suggests the cause of this precipitation is likely a process other than field-line scattering, such as wave-particle interactions, and therefore unrelated to the IB. Panel 6 shows the net precipitating electron energy flux, integrated over energy and solid angle within the loss cone. This is used to estimate both the relative and total amounts of average precipitating energy flux associated with the isotropy boundary. Panels 7-10 display the pitch-angle spectra for select energy channel ranges, alongside horizontal bars indicating the local bounce-loss/anti-loss cones. Panel 11 shows the IGRF magnetic field at the spacecraft location, for reference.\\

\subsection{Statistical characterization}
To obtain the statistical results of the study, we performed the preceding type of analysis on each ELFIN QSZ data collection. We then manually inspected the events for the presence of an electron isotropy boundary. To assess whether an IB was present we computed the flux isotropy ratio $R_I=J_{prec}/J_{perp}$ and checked for energy-latitude dispersion signatures consistent with the field-line curvature scattering mechanism (e.g. as in Fig. 1). In order for a science zone to be marked as containing an IB crossing, and to distinguish it from other potential precipitation processes, the candidate crossings were required to satisfy several criteria: 1) a poleward transition from anisotropy to isotropy (defined as $R_I \geq 0.6$) for energy channels with non-zero counts, followed by persistently isotropic flux $(R_I\sim 1)$ for at least 50\% of the subsequent non-zero poleward data samples; 2) at least three energy channels satisfying condition 1), and thus usable towards identification of an IB dispersion signature; 3) a negative data-fitted energy-latitude dispersion slope, one consistent with higher-energy electrons becoming isotropized at lower latitudes (or, at most, at apparently the same latitudes given the finite spin-period time-resolution of ELFIN measurements). To limit the effects of low-counting statistics, the quantity $R_I$ was assumed to be 0 for cases in which the measurement uncertainty was more than a factor of 2. Additionally, we rejected IBs with equatorial footpoints residing within 3 Earth radii (or outside of) a model magnetopause \cite{Fairfield1971}. This was done to eliminate false IB-like particle signatures which may result from scattering at the magnetospheric boundary (particularly on the dayside and flanks). For cases which satisfy all of the preceding criteria, the event was marked as containing an IB crossing. \\

We next determined the occurrence rate distribution of IB crossings in our database of ELFIN QSZ collections versus MLT, magnetic latitude (MLAT), L-shell (IGRF or T89-based), and geomagnetic activity (based nominally on AE, and including Dst and Kp for characterization of IB-associated energy fluxes). We also binned likewise the IB-associated (latitude-averaged over the PS2ORB interface), integral (over energies $\geq$50 keV) electron precipitating energy flux. This average was presumed to be dominated by field-line curvature scattering. To obtain the average flux precipitating due to field-line scattering over the PS2ORB interface, we used the following operational definition of the PS2ORB interface region: The equatorward-most PS2ORB interface latitude was taken to be the IB for the maximum observable isotropic electron energy, while the poleward-most PS2ORB interface latitude was taken to be the omni-directional flux cutoff in $\geq$300 keV electrons poleward of its IB (see the Results Section for rationale). For cases in which the 300 keV electron channel was not present (or its IB could not be clearly identified), the next-closest energy channel cutoff was used to mark the PS2ORB poleward-most latitude. From these events we also determined the minimum and maximum electron energies of the IB crossing dispersion, alongside the average energy-latitude slope of the dispersion. We lastly computed the latitudinal width of the combined IB plus PS2ORB interface region, which provides an estimate of the spatial extent of the equatorial FLCS-dominated region at the collection time.

\section{Results}
\subsection{Occurrence, spatial distribution, and dispersion of electron IBs}
To provide an initial picture of the spatial distribution of electron IBs in our dataset, the equatorial footpoints corresponding to the minimum and maximum L-shell of field-line traces bounding the IB dispersion region (based on T89) are shown in Figure 3 as a scatter plot, projected in the equatorial GSM xy-plane. The points marked $L_{min}$ represent the most Earthward portions of the IB crossings (typically corresponding to the highest-present electron energy IB; variable from event to event), and the point marked $L_{max}$ similarly represents the furthest portion of the IB dispersion in the crossings (almost always corresponding to the 50 keV electron IB). The static reference-magnetopause based on \cite{Fairfield1971} is shown (solid) alongside the 3 earth radius cutoff (dashed) used to reject false IB signatures, which are marked in silver and black (counted as non-IB but otherwise valid events). It is evident that projected isotropy boundary locations span a wide range of MLTs and mapped equatorial distances throughout much of the nightside magnetosphere.\\

Figure 4 shows the electron IB spatial distribution. The top panel depicts the histogram of isotropy boundary occurrence (absolute and normalized) versus MLT in our database, binned with a bin size $\Delta MLT = 1$ hour. Blue bars denote the number distribution of all QSZ events in our database, orange bars depict the number of events with an IB (both referring to the absolute numbers on the left vertical axis), and red line is the normalized occurrence rate of IB crossings versus MLT in the dataset (corresponding to the right vertical axis), computed as the ratio of the orange to blue values. A running average with $\pm$1-hour MLT window was applied to the data, to improve statistics and reduce uncertainties, e.g., in field-line mappings. \\

We find that IB crossings occur in up to 90\% of events near midnight, and exhibit a sustained occurrence rate of $\geq$80\% between 21 to 01 MLT. Their occurrence rate gradually declines towards near-zero at pre-dawn (7 MLT) and near dusk (16 MLT). Interestingly, the peak IB occurrence is not centered at midnight but at pre-midnight, around 22 MLT. (We note that the apparent peaks in event number around 01 and 15 MLT are due to the reduced ELFIN data availability in the first two years of the mission used here; i.e., in the 2019-2020 period.) The associated spatial distribution of these IB events binned by L-shell ($\Delta L = 1$) and magnetic latitude ($\Delta$MLAT = 1$^\circ$) using the IGRF model are shown in Fig. 4 middle and bottom, respectively. IGRF alone, not T89, was used to compute the distribution in this case to provide a static reference independent of the behavior of the magnetic field far from earth, with the caveat that this will typically underestimate the true L-shell and latitude mappings at the nightside as distances from Earth increase, and for different solar wind driving. We see that electron IBs span a wide range of L-shells (4 to beyond 12), and are most commonly found to span L-shells 6-8, typically corresponding to the tail-dipole transition region, or PS2ORB interface region. As anticipated from the L-shell to $|$MLAT$|$ equivalence in mapping, a similar trend is observed in magnetic latitude: a broad range of occurrences between $60^\circ$ to $74^\circ$, with peaks around $66^\circ$-$70^\circ$. Additionally, we checked the occurrence of IB crossings versus the geomagnetic activity indices AE, Dst, and Kp (not shown), and found them to not vary appreciably from the overall MLT distribution described here.

In these events, we also characterized the minimum and maximum electron energies of the observed dispersion in the IB crossings, alongside the latitudinal spread and dispersive slope (``sharpness''), versus MLT and activity. Given that the IB generation mechanism is based on the background magnetic field configuration and the local availability of energetic particles (e.g. injections), it is reasonable to expect that the level of geomagnetic activity and solar driving at the time of (and preceding) the crossing would affect the electron energies appearing in the IB, as well as the latitudinal onset and sharpness of isotropic dispersion. In order to provide an assessment of this activity-dependent behavior, we separated the dataset into two categories: quiet-time intervals with 1 hour average AE less than 200 nT, and non-quiet intervals with 1 hour average AE above 200 nT. This choice of hourly-average AE split the dataset roughly in half and aimed to emphasize high-latitude geoeffectivity while avoiding complications of the potential time-delayed nature of other indices (such as Dst at storm time).\\

Figure 5 displays the activity-dependent minimum and maximum observed IB electron energies versus MLT, where solid points represent the mean values and error bars represent the occurrence-weighted standard deviation about the mean. The top panel shows the minimum and maximum electron energies appearing in the dispersion of IB crossings within the ELFIN EPD instrument sensitivity and energy limits. For low activity (AE$<$200 nT) the electron energies span a typical maximum of 700-800 keV, peaking at MLTs near midnight, and trailing toward a lower average maximum in the 300-400 keV range near dawk and dusk. For higher activity (AE$>$200 nT), the maximum energies rise dramatically to a mean of 2 MeV, and are shifted in local time to pre-midnight (peaking around 22 MLT), with the maximum energies falling to 800-900 keV at MLTs approaching dawn and dusk. The apparent shift in peak energy to the pre-midnight during active times is consistent with the idea of the FLCS source being supplied by frequent (possibly continual) local energetic flux injections in the tail at these MLTs \cite{Gabrielse2014,Liu2016}, which the background field rapidly isotropizes. At both quiet and non-quiet times the low-energy minimum of the IB is almost always the lowest energy channel resolvable by the ELFIN EPD ($\sim$62 keV mean). This suggests that the IB likely extends to lower energies than ELFIN can resolve, and that such fluxes are highly available in the tail for FLCS at the rates reflected by IB occurrence in each MLT sector.\\

Using this activity criterion, we also determined the MLT-binned latitudinal bounds and sharpness of the energy-latitude dispersion in the IB crossings. The center row panels in Figure 5 show the IGRF L-shell distribution (left) and magnetic latitudes (right) bounding the onset in isotropic dispersion across all resolved energies. The quantities denoted ``min'' reflect the most Earthward (i.e. lowest latitude, often highest-energy) portion of the dispersion while the ``max'' values mark the more distant location at which the lowest energy channel is first observed to become isotropic. The results show that at low activity, there is a symmetric bowl-shape distribution with global minimum around 22 MLT with $L_{min}\sim 7.1$, $L_{max}\sim 7.3$ and $MLAT_{min}\sim 67.6^\circ$, $MLAT_{max}\sim 67.9^\circ$, rising to a maximum of $L_{min}\sim 8.6$, $L_{max}\sim 8.9$ and $MLAT_{min}\sim 69.8^\circ$, $MLAT_{max}\sim 70.0^\circ$ at 5-6 MLT. Interestingly at both activity levels the mean latitudinal width (max minus min) of the dispersion in the crossings is nearly constant across MLT, suggesting a highly persistent supply of injected particles of appropriate energies (top panel) and appropriate background field configuration across many local times on the nightside, such that they are repeatably affected by FLCS. At higher activity, IB dispersion is found consistently at lower latitudes across all MLTs, with the emergence of a break from the symmetric bowl-shape distribution around 22 MLT. This feature is again consistent with the appearance of energetic electron injections at lower latitudes, e.g. during substorms. Rather than being localized to a single MLT, the active time latitudinal onset minima are instead spread between 20-23 MLT, taking on values of $L_{min}\sim 6.3$, $L_{max}\sim 6.5$ and $MLAT_{min}\sim 65.9^\circ$, $MLAT_{max}\sim 66.5^\circ$ and maxima around 4-5 MLT of $L_{min}\sim 7.6$, $L_{max}\sim 8.0$ and $MLAT_{min}\sim 68.4^\circ$, $MLAT_{max}\sim 68.8^\circ$. We note that MLT sectors with fewer than 5 IB events are not shown, which is the reason there are no data points for MLT 5-6 at active time versus quiet time.\\

Based on the observed energy and latitudal ranges of the IB crossings, we finally computed the linear energy-latitude dispersion slopes $dL/dE$ and $dMLAT/dE$ versus activity and MLT in the top and center panels. The bottom panel shows the dispersion slope in terms of L-shell (left) and MLAT (right). The data reveal that during quiet intervals the MLT-based distribution is similarly bowl-shaped with a minimum slope (i.e. steepest latitudinal change in energy) around midnight. As with the energies and latitudes, the dispersion slope extrema shift to the pre-midnight at active times, with a universal trend toward apparently sharper IBs. Such values provide insight into both the equatorial profile of the magnetic field under these conditions and the typical energies available from injections there. We note again that while IGRF provides a practical activity-independent spatial reference, the reported IB latitudes and slopes would be quite different (especially away from midnight) versus models containing magnetospheric currents, such as the magnetotail and the ring current (e.g. T89).

\subsection{Precipitation associated with electron IBs}
Next, we seek to quantify the amount and distribution of precipitating energy flux associated with electron isotropy boundaries. The isotropy boundary for each energy is by definition an instantaneous transition latitude (separating isotropy from anisotropy), and consequently does not capture the finite spatial extent of precipitation associated with the FLCS isotropization, which is expected to extend into adjacent poleward latitudes (as seen in Fig. 2). This prompts us to use an operational definition of the poleward extent of the near-isotropic precipitation associated with the IB. To achieve this, we rely on the fact that the inner edge of the plasma sheet proper (for which particle processes other than FLCS may become dominant) can be regarded as a proxy for the poleward edge in FLCS-dominated fluxes. To identify this edge in the data, we relied on the fact that the plasma sheet typically possesses a high-energy cutoff in electron fluxes at both quiet and active times \cite{Christon1989,Christon1991}. By statistically determining the most common maximum electron energy at which omni-directional fluxes experience a sustained drop-out at latitudes poleward of IBs, we were able to define an operational definition of the poleward edge of the IB-associated precipitation region, otherwise referred to here the plasma sheet-to-outer radiation belt interface region (``PS2ORB''). We note that the specific upper plasma sheet cutoff energy found by this method depends on the sensitivity and resolution of the instrument used to measure it, and thus had to be determined using the ELFIN dataset (akin to that performed with ISEE observations in Christon et al.).\\

Figure 6 explores the energy-latitude dependence of omni-directional electron flux cutoffs poleward of IBs in the ELFIN dataset, which are used to define the operational outer edge of the PS2ORB region. The top panel depicts as a function of energy the mean and median IGRF magnetic latitude of the lowest latitude (highest energy) portion of all IB events for comparison with the poleward cutoff in omni-directional electron fluxes. For energies $\geq$300 keV, there is a near-constant $\sim$$1^\circ$ separation in average magnetic latitude between the lowest latitude portion of the IB crossings and of the omni-directional flux dropouts. For energies $<$300 keV, this difference begins to grow rapidly to $>$$3^\circ$, suggesting that a separate electron population from the plasma sheet has been encountered. The bottom panel repeats this analysis instead using the event-wise difference between the lowest latitude IB and the poleward omni-flux cutoff. This value is also around $1^\circ$ at $\geq$300 keV but suddenly increases with decreasing energy, providing additional confirmation that the latitude of flux cutoff at 300 keV in each specific event is a reasonable proxy for the location of the poleward boundary of the IB FLCS-dominated precipitation region. We thus defined the poleward bounding latitude of the PS2ORB region as the latitude beyond the IB at which $\geq$300 keV omni-directional fluxes first drop out. Using this definition, the bottom panel shows the mean latitudinal width $\Delta \theta$ of the PS2ORB region observed in the dataset versus local time during quiet and non-quiet intervals. The data reveal that at quiet intervals, the latitudinal width of the FLCS-dominated region is typically between 1-1.5$^\circ$, rising to 2-3$^\circ$ at non-quiet times. This is consistent at active times with the IB/FLCS source having greater access to source particles to be scattered (such as from injections and enhanced electric fields in the magnetotail), as well as from modified equatorial background magnetic field properties (such as during intervals of magnetospheric compression), which can further shift and extend the FLCS-dominated region in latitude.

Using the above operational definition for the latitudinal bounds of the PS2ORB interface region, we computed the average precipitating energy flux within IB-associated latitudes from isotropic $\geq$50 keV electrons. We also computed the net energy flux integrated over all latitudes including and poleward of the PS2ORB region, and for the entire ELFIN science zone (from all available latitudes -- typically spanning $55^\circ$ to $80^\circ$). This was done to provide a comparison of the relative magnitude of IB-associated energy flux versus that which is sourced by other ELFIN-observable regions, such as the outer radiation belts, electron plasma sheet, and the polar cap. To compute the energy flux, we integrated the precipitating electron distributions over energy and solid angle (pitch-angle and gyrophase), and averaged over the number of data samples the spacecraft spent in each region in each event. Figure 7 shows the distribution of energy flux in each of these latitudinal ranges, including the entire science zone (blue), the IB crossing and regions poleward of it (orange), and the IB/FLCS-dominated PS2ORB interface region (green). We immediately see that the PS2ORB interface region exhibits the highest average and maximum of the three categories. Comparing the green and orange bars also reveals that the precipitation is confined in latitude to the localized PS2ORB interface region adjacent to electron IBs, rather than being evenly distributed over all poleward latitudes (e.g. in the plasma sheet and polar cap).\\

To further investigate the occurrence and significance of IB-associated precipitation, we computed the local time distribution of latitude-averaged precipitating energy flux within the PS2ORB interface, as well as the fraction of the net energy flux (over the entire ELFIN science zone) contributed by the PS2ORB region. To assess how these quantities varied with geomagnetic activity, they were also binned by the activity indices AE, Dst, and Kp, taken independently of local time. We computed the total power deposition as the product of the crossing time and average flux. This is because as a polar-orbiting ELFIN satellite moves at roughly constant velocity across different latitudes in space (approximately at the same longitude), the time-average of the flux represents a line-integral over the latitudinal spatial dimension. The resultant quantity is power deposited over an ionospheric swath along the satellite track per unit cross-track distance (in ergs/s/cm or, re-scaled, in Watts/km). Thus the time-average represents average power deposition at the ionosphere at a given MLT, either over the entire science zone crossing, or over the more limited PS2ORB region. Ratios of time-average total precipitation in the PS2ORB region over that in the entire science zone represent the fractional total energy per unit time, or fractional power, in PS2ORB relative to the whole science zone. Figure 8 captures the total fractional power (top) and average electron energy flux (bottom) of the precipitation associated with the PS2ORB interface region (the IB FLCS source). For each MLT sector (horizontal axis) the value in color (see color bar) shows the fraction of the IB events possessing a value greater or equal than shown vertical axis (i.e., cumulative proability). For example, we can see in the top panel than 0.15 (or 15\%) of PS2ORB events (orange color in color bar) contribute, near midnight, $>$0.4 ($>$40\%) of the total precipitating power within their individual science zones, and we can see in the bottom panel that 0.1 (or 10\%) of PS2ORB events (yellow-green color) have, near midnight, an average energy flux $\geq$10$^{-1}$ erg/cm$^2$-s. Note that here, as with the IB occurrence vs MLT distribution, we performed a three-hour MLT average on these values (central value plus and minus 1 hour). Pink lines in the two panels represent means of relative precipitating power (top panel) and average precipitating flux (bottom panel).\\

To summarize key features of these results, we see (top panel, pink line) that within $\pm$4 hours of pre-midnight, the PS2ORB interface region precipitation accounts for 0.1-0.2 (10\%-20\%) of the total precipitating power, on average. While this is a small value, on average, the cumulative probability distribution (top panel, color plot) shows that precipitating power $>$0.5 ($>$50\%) of the total within a science zone (vertical axis value: 0.5) occur, near midnight, 15\% of the time (yellow-orange colors: 0.15 in vertical color bar). In extreme cases, near 22 MLT, up to $\sim$100\% of the power can be delivered by PS2ORB (albeit more rarely). The average total precipitating energy fluxes (bottom panel) are found to span several orders of magnitude across all MLTs. Interestingly, the peak relative and total amounts of precipitating energy flux again occur in the pre-midnight sector around 22 MLT. As alluded to previously, this is likely due to preferential current sheet thinning at pre-midnight, the related proximity of the cross-tail current to Earth and the preponderance of substorm onset, and injections at that location.\\

The results from Fig. 8 additionally allow for an estimate of the total global atmospheric power input across all local times and latitudes from IB-associated $\geq$50 keV electron precipitation. To determine the average global power input across the dataset, we used the IB MLT and latitudinal occurrence rates from Fig. 4 combined with the average PS2ORB latitudinal extent from Fig. 7. Using the pink mean line from Fig. 8, we add up the occurrence-weighted total energy flux in each 1 hour (15$^\circ$ longitude) MLT sector scaled by the effective area of the PS2ORB projected onto the atmosphere at the IB latitude, assuming 1$^\circ$ in latitude equates to 111 km projected, and that both hemispheres contribute equally. This results in a typical total atmospheric power deposition of around 10 MW at any given time. However, in the most extreme (and rare) cases where energy fluxes approach 1 erg/cm$^2$-s across several several hours in MLT, the total atmospheric power input from IB-associated $\geq$50 keV electron precipitation can exceed 1 GW---possibly exceeding the input from auroral sources, and thus likely playing an important role in ionospheric processes.

\subsection{Energy flux variations with geomagnetic activity}
Lastly, we explored the dependence of precipitation on geomagnetic activity. Towards that goal we computed the distribution of relative precipitating power and total energy flux in the PS2ORB interface region (IB-associated) versus geomagnetic activity indices AE, Dst, and Kp. Figure 9 presents these results. The top row corresponds to the PS2ORB interface region precipitating power relative to the total power in the science zone (akin to Fig. 8 top) while the bottom row contains the average precipitating energy flux (akin to Fig. 8 bottom). The AE values ($\Delta AE = 100$ nT) were three-hour averaged leading into the collection interval in order to capture longer-term activity trends rather than individual short-term activity phases (e.g. as in substorms). The final bin is integral covering the cases of AE$\geq$600 nT. The trend is for both the relative power and total energy flux of PS2ORB precipitation to increase with AE (for AE increase from 0 nT to 600 nT the relative power increases by 30\% and the total energy flux by several orders of magnitude). A similar increase in PS2ORB precipitation is evident for relative power and for total energy flux as Dst decreases from 0 nT (quiet times) to $<$-30 nT (storm-like times), as shown in Figure 9 in increments of $\Delta Dst = 10$ nT. In this case there is also a rise for positive Dst values, which may correspond to effects associated with Storm Sudden Commencements (SSCs). We observed a similar relationship between Kp ($\Delta Kp = 1$) and the precipitating fluxes that was seen for AE: both the relative power and total energy flux increase nearly monotonically with this index. There are also changes in the slope of the relative intensities at Kp 2 and 5, indicating that on average, magnetospheric dynamics may exhibit an increased preference for FLCS over other processes under these conditions.\\ 

\section{Summary and discussion}
Using $\sim$1900 ELFIN-A and -B science zone collections from years 2019 and 2020 we have characterized the occurrence and spatial distribution of electron isotropy boundaries and associated electron energies and precipitation from $\geq$50 keV electrons due to magnetic field-line curvature scattering. We examined the spatial distributions in MLT, L-shell, and magnetic latitude and the dependence of precipitating power and energy flux on geomagnetic activity indices AE, Dst, and Kp. We found that electron IBs are present under all activity levels over nightside local times, from dusk through midnight to dawn. They have $\sim$90\% peak occurrence rate around 22 MLT, remaining above 80\% occurrence rate between MLTs of 21 to 01. The most common latitudes associated with IB energy-latitude dispersion are between 66$^\circ$-68$^\circ$ (IGRF), or L-shells between 6 and 8, with a total observed latitudinal range between $60^\circ$ and $74^\circ$, or $L\gtrsim 3.5$. The latitude-averaged precipitating energy flux associated with IBs, as well as the contribution to the total high-latitude precipitation power also peaks around 22 MLT, and has an MLT distribution similar to that of the IB spatial occurrence rate. The peaks in electron IB occurrence in MLT and magnetic latitude and in precipitation energy are statistically collocated with substorm onset MLT and latitude. 

The evolution of electron IB latitude and peak energy with activity is also consistent with the equatorward evolution of the edge of the oval during active times. This suggests that geomagnetic activity is conducive to the formation of better-defined IBs, which implies more intense fluxes at the tail-dipole transition region. We interpret this as due to the availability of intense fluxes over broad energies from 10s to 100s of keV (and often $>$1 MeV) of electrons injected to that region by reconnection outflows, as in \citeA{Gabrielse2014}. The maximum average IB-associated energy flux observed was on the order of $\sim$1 erg/cm$^2$-s but varies considerably spanning four orders of magnitude depending on geomagnetic activity. The IB/FLCS source contributes, on average, between 10-20\% of the total nightside high latitude power, but contributes more than 50\% of the total around 20\% of the time (in rare cases even approaching 100\% of the total). The total global atmospheric power input from the IB/FLCS source is on average around 10 MW, but can exceed 1 GW in the most extreme cases. Both the total intensity and the relative contribution to high-latitude precipitation of the equatorial FLCS source are seen to increase with increasing AE, $|$Dst$|$, and Kp, suggesting that IBs may play an important role in high-latitude energetic electron precipitation at active times.

Our study establishes a baseline for further investigations of the effects of energetic electron precipitation from IB-associated field-line scattering for magnetospheric and ionospheric processes. While we do not expect the precipitating power from the IB source to typically out-compete the total power from lower-energy ($<$50 keV), higher number flux auroral sources at higher latitudes (e.g. as in \citeA{Newell2009}), more than 50\% of the IB events in our database contained $\geq$500 keV precipitating electrons. This indicates that on average the IB-associated source will tend to penetrate much deeper into the atmosphere than 10s of keV particles, and it is therefore worth considering in models of ionospheric conductivity and chemical reactivity at lower altitudes. This also suggests that models of energetic particle transport from the plasma sheet to the outer radiation belts should also incorporate these results for a proper accounting of magnetospheric losses. The electron IB properties are all found to be positively correlated with geomagnetic activity as inferred from splitting the dataset about AE=200 nT. Further investigation is needed to determine the causal relationship between solar wind driving and magnetospheric responses that result in IB-related precipitation. These studies could be further augmented by conjunctions with equatorial spacecraft which may provide simultaneous equatorial observations of the magnetic topology and electron populations, while also reducing field-line mapping uncertainties.\\

Similar to the proton IBs (e.g. as in \citeA{Newell1998}), electron IBs exhibit peak pre-midnight occurrence rate and precipitation intensity around 22 MLT. This is presumably due to the proximity of the cross-tail current to Earth in that region due to its magnitude being stronger at dusk than at dawn \cite{Lu2018}. This results in a tail-dipole transition region closer to Earth at dusk than at dawn, which can be associated with sharper field-line curvature. Moreover, we note that the highest energy portion of electron IBs is sometimes observed to occur in the vicinity of characteristic signatures of electromagnetic ion cyclotron (EMIC) wave scattering in ELFIN data. Such signatures are most common at pre-midnight (where EMIC waves peak in occurrence rate). This suggests that occasionally both mechanisms can act at the same location, and may, in fact, share common driving: magnetotail injections. Beyond the technique used to identify IB dispersion, we did not attempt to separate the two processes in our analysis, but we note that in future studies it is possible to distinguish the two thanks to the characteristic narrow-banded energy and lower-latitude extent of EMIC precipitation.\\

Around 5\% of events in our dataset exhibit unusually intense IB-like wide-energy isotropic electron precipitation signatures at either very low latitudes (appearing as low as L$\sim$4 near the plasmapause), or over a highly extended latitudinal range (exceeding $10^{\circ}$ in apparent poleward extent), corresponding to significantly disturbed magnetospheric conditions during storms or strong substorms. We excluded such events from our database because they did not exhibit the characteristic energy-latitude dispersion of the isotropy boundary assumed in the study. Nonetheless, their properties suggest that wave-particle scattering sources alone may not sufficiently explain them, leaving FLCS as a potential cause. One possibility is that these structures correspond to an active-time type of isotropy boundary in which the equatorial magnetic field is varying over FLCS timescales, possibly accompanied by rapid particle energization in vicinity of the scattering region. Such signatures may be compatible with those of local tail $B_z$ minimum as reported e.g. by \citeA{Sergeev2018}, or of active tail reconfiguration, which would appreciably vary the critical FLCS scattering threshold $\kappa_{cr}^2$ and field-line mappings in real-time. Additionally, during periods of moderate activity, strongly populated IB-like structures can also sometimes be seen at dawn/dusk. A possible explanation may be that ULF waves of appropriate frequency and comparable strength to the equatorial magnetic field may allow for electron isotropization far away from midnight. This would present an alternative mechanism to magnetopause scattering at the flanks, effectively expanding the IB local time occurrence distribution to events that were filtered out by the 3 earth radius magnetospheric boundary culling criterion in our analysis, as well as substantially increasing the maximum amount of precipitating energy flux contributed by IBs/FLCS. Further simulation and observational work will be helpful in addressing these and related questions.\\

In addition to its relevance for electron loss and atmospheric energy deposition, our study also has implications for improvements in magnetic field modeling. Such models are still far from encompassing dynamical variations of the magnetotail in space and time. By forming an observational dataset of isotropy boundary properties of the type used here, magnetic models can be further constrained (e.g. as alluded in \citeA{Sergeev1993}), either by fine-tuning their existing parameters (such as cross-tail current sheet thickness and Earthward-most extent), or by direct value sampling in modern assimilative models. This method becomes especially powerful when combined with equatorial measurements of the magnetic field for mapping, constituting an avenue ripe for future effort. We note that in this work, predictions based on magnetic field models were only used as a qualitative estimate of the expected isotropy boundary properties.\\

A final clarification concerns the extended isotropic precipitation region associated with and poleward of the IBs, which we referred to as the ``electron plasma sheet-to-outer radiation belt'', PS2ORB. Using ELFIN data alone at a single local time, it is not feasible to ascertain a definite instantaneous inner edge to the electron plasma sheet, nor to infer the location of the last-closed drift shells of the outer belt. We regard this region as the transition in global background magnetic field from the outer to the inner magnetosphere in which freshly injected plasma sheet electrons acquire sufficient energy to be visible above the highest plasma sheet energies ($>$300 keV) and for which freshly injected, trapped or quasi-trapped outer belt electrons curvature scatter on timescales faster than a drift period. Both populations then manage to precipitate with high intensities prior to reaching the low-latitude boundary layer or the magnetopause boundary (meeting the phenomenological definition of a transition separating the plasma sheet from outer belt). We envision that precipitation from FLCS in this region out-competes other processes (as evidenced by the energy-latitude distribution of flux cutoffs that dominates over other wave-associated particle spectral shapes, e.g., due to whistler-mode chorus or EMIC waves). This assumption can benefit from ELFIN comparison with in-situ equatorial measurements of electron flux, wave power and magnetic field strengths, as well as from data-constrained adaptive magnetic field models.

\section{Open Research}
The ELFIN data and software used in this study are part of the SPEDAS framework \cite{Angelopoulos2019}, which is freely available to the public at the following url: \url{http://spedas.org/wiki/index.php?title=Main_Page} 






\acknowledgments
This work has been supported by NASA awards NNX14AN68G, 80NSSC19K1439, and NSF grants AGS-1242918 and AGS-2021749. We are grateful to NASA's CubeSat Launch Initiative for ELFIN's successful launch in the desired orbits. We acknowledge early support of ELFIN project by the AFOSR, under its University Nanosat Program, UNP-8 project, contract FA9453-12-D-0285, and by the California Space Grant program. We acknowledge critical contributions of numerous ELFIN undergraduate student interns and volunteers.

\newpage
\bibliography{bib_ib_paper}

\begin{figure}
\noindent\includegraphics[width=\textwidth]{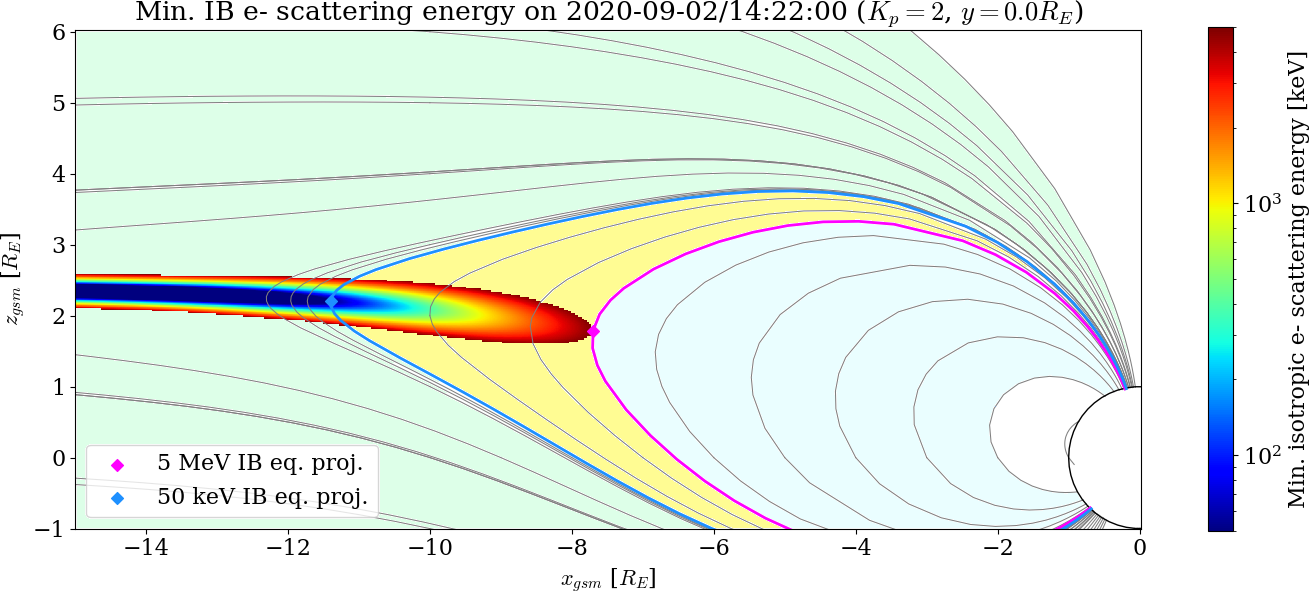}
\caption{Model-based spatial profile (GSM xz-cut) on 2020-09-02/14:22:00 of equatorial source locations corresponding to minimum electron kinetic energies for isotropization by field-line curvature scattering, resulting in an isotropy boundary observed at LEO. The minimum required energy for scattering is observed to be lowest at the center of the cross-tail current. The pink and blue curves depict the field-line mappings for the 5 MeV and 50 keV IBs, respectively. Clear energy-latitude dispersion is observed in the IB location for a given energy, with more energetic particles isotropized closer to Earth. The region shaded in yellow shows the set of field-lines mapping into the equatorial energy-latitude dispersion region, while the mint and light blue colors show the poleward and equatorward latitudinal extent ELFIN can typically observe. The critical field geometry parameter $\kappa^2_{cr}$ was taken to be 8.}
\label{Figure 1.}
\end{figure}

\begin{figure}
\vspace{-1cm}
\noindent\includegraphics[width=\textwidth]{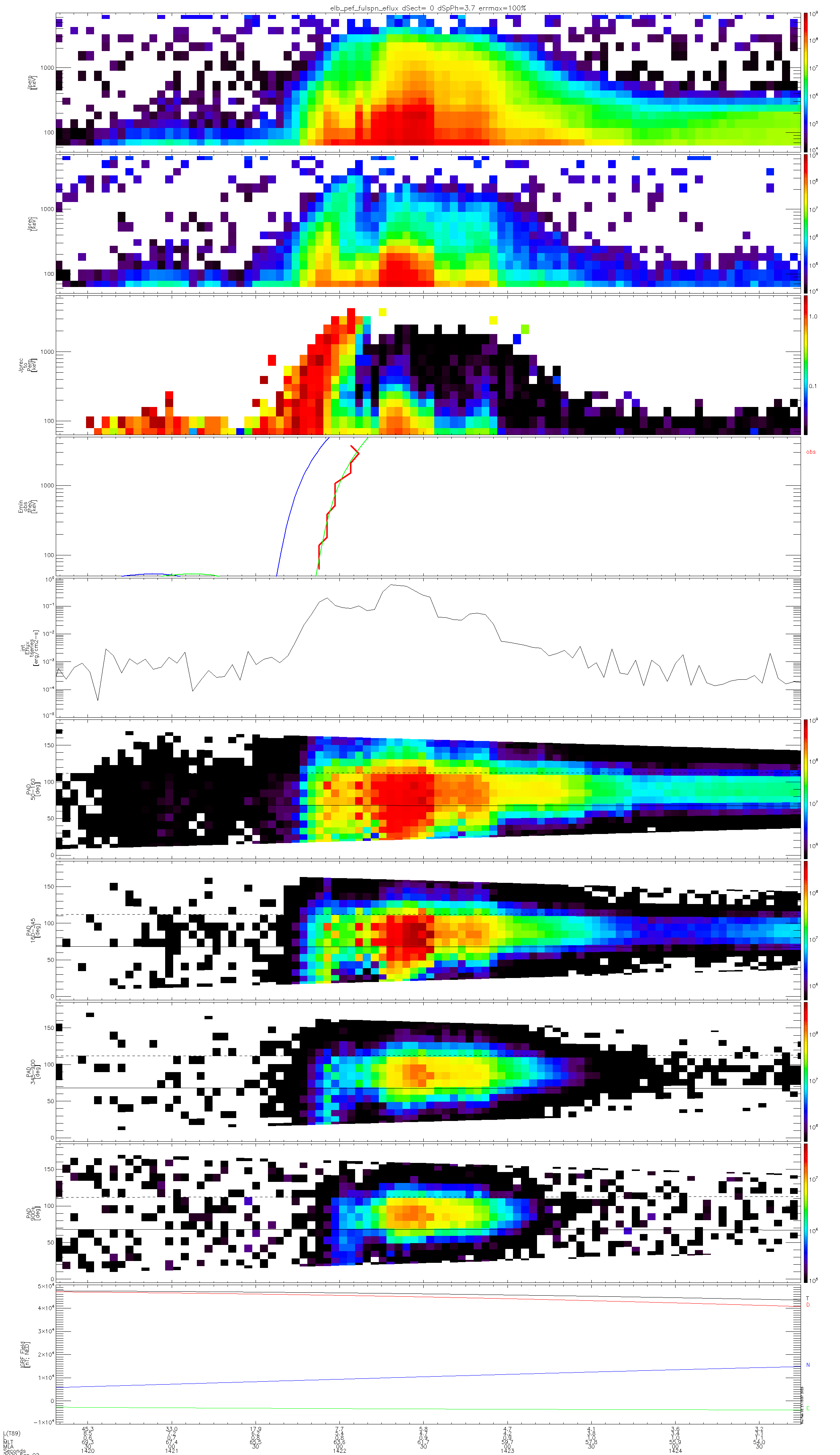}
\caption{Example isotropy boundary crossing observed by the ELFIN-B on 2020-09-02. The spacecraft began in the northern polar cap and moved southward toward the equator, crossing first plasma sheet and then outer radiation belt field lines. The top two panels show perpendicular (locally-mirroring) and precipitating electron energy spectra, respectively. The third panel shows the instantaneous flux isotropy ratio spectrogram $R_I=J_{prec}/J_{perp}$ computed using data from Panels 1 and 2. This ratio exhibits a clear (abrupt) poleward transition (backwards in time) from anisotropy (low ratio) to isotropy (high ratio) at a given energy, at $\sim$14:22 UT. The latitude of that transition is the IB for that energy. The dispersion in energy versus IB (latitude) is clear and suggests that the 50 keV to 4 MeV electron IBs were crossed at around that time.}
\label{Figure 2. }
\end{figure}

\begin{figure}
\noindent\includegraphics[width=\textwidth]{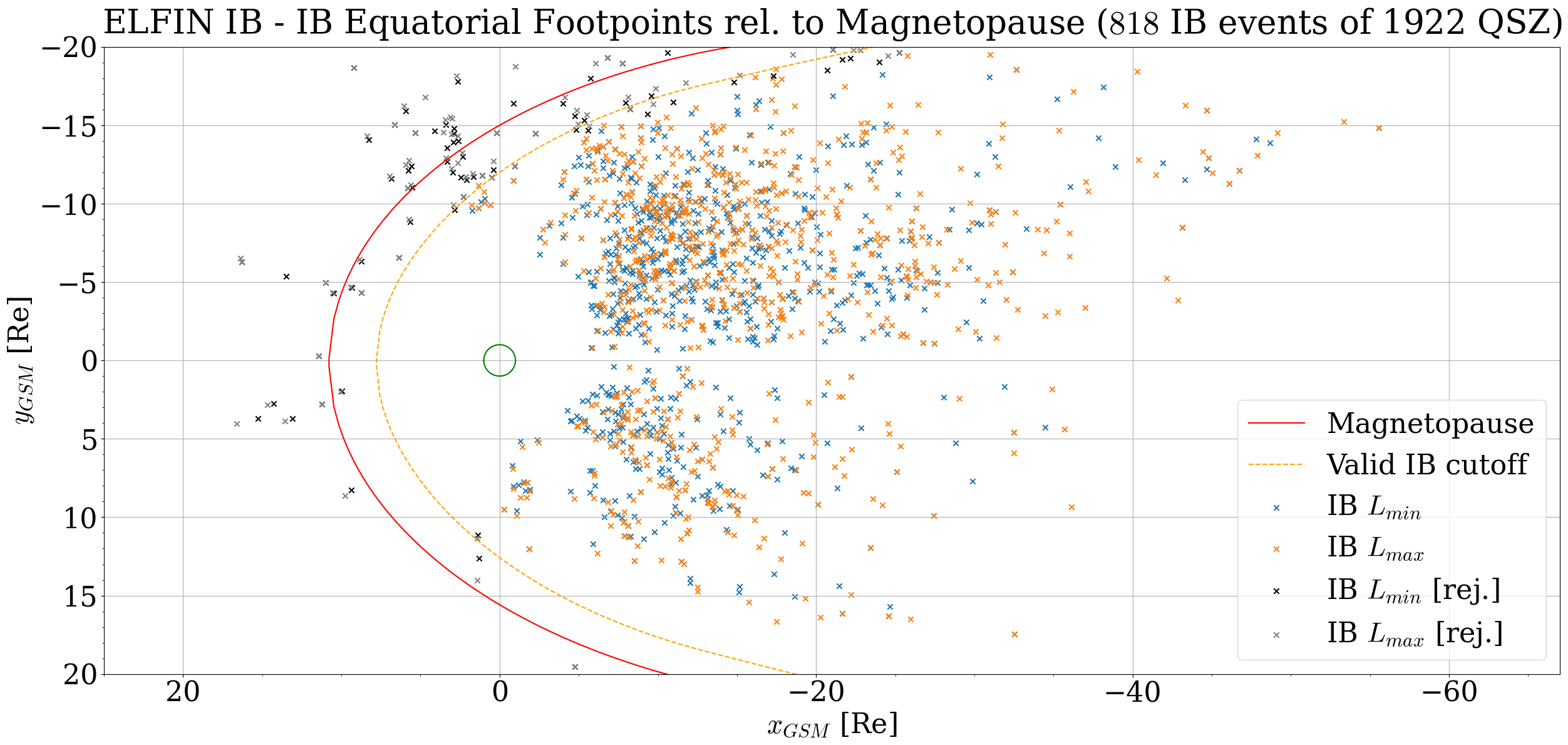}
\caption{Spatial distribution of isotropy boundary locations in our dataset, observed by ELFIN and magnetically mapped to the equator. Points marked in blue represent the closest Earthward footpoint of each observed IB crossing (nominally corresponding to the highest energy observed to have isotropized electrons). Orange points represent the furthest footpoint in the IB dispersion signature (almost always representing the onset of 50 keV electron isotropy -- the lowest energy ELFIN can measure). Points marked in gray and black represent the same IB dispersion signatures as blue and orange, respectively, but are too close to the magnetopause (solid red line). Those crossings (both gray and black) have been rejected as likely being of magnetopause origin, based on the criterion that the 50 keV IB (gray point) is farther from Earth than 3 Earth radii inside of the magnetopause (dashed yellow line).}
\label{Figure 3. }
\end{figure}

\begin{figure}
\noindent\includegraphics[width=\textwidth]{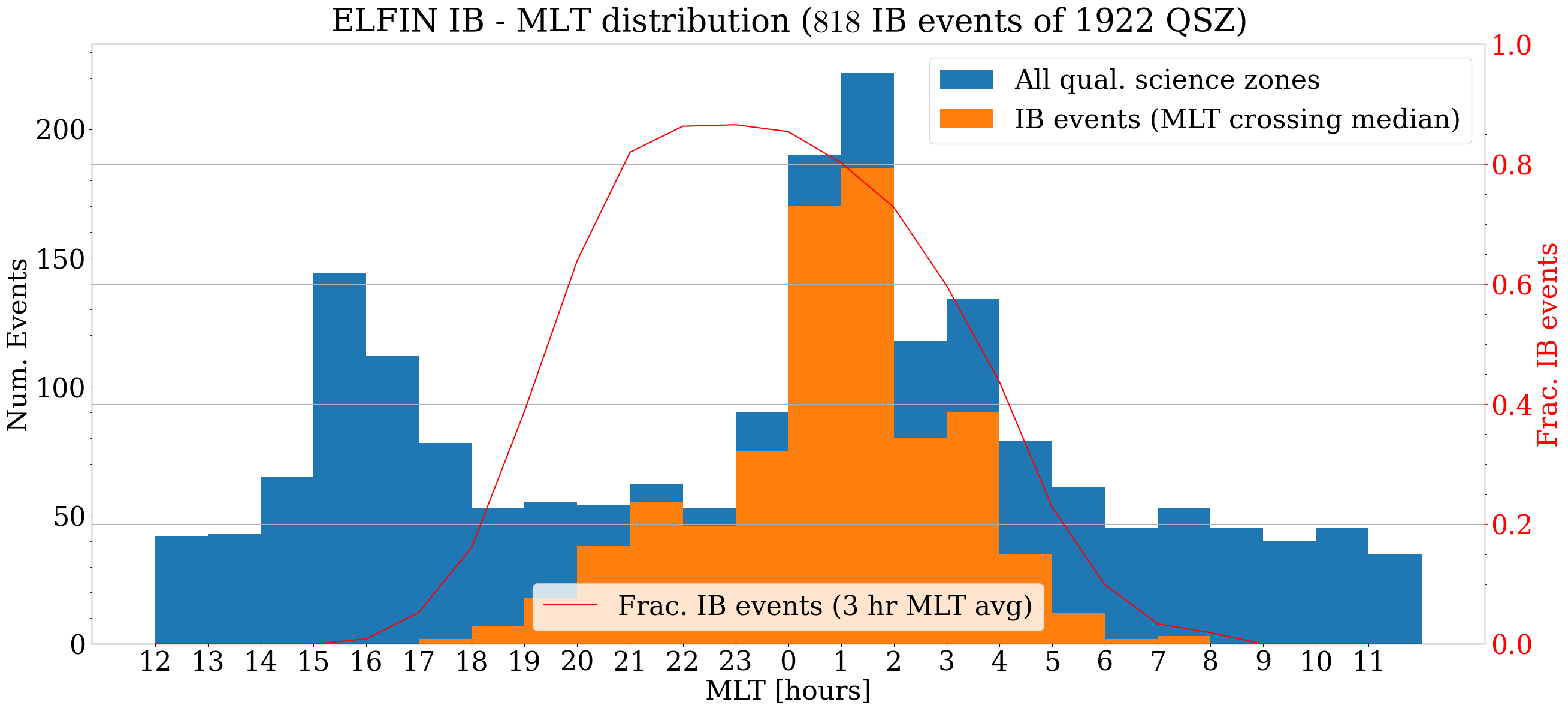}\\
\noindent\includegraphics[width=\textwidth]{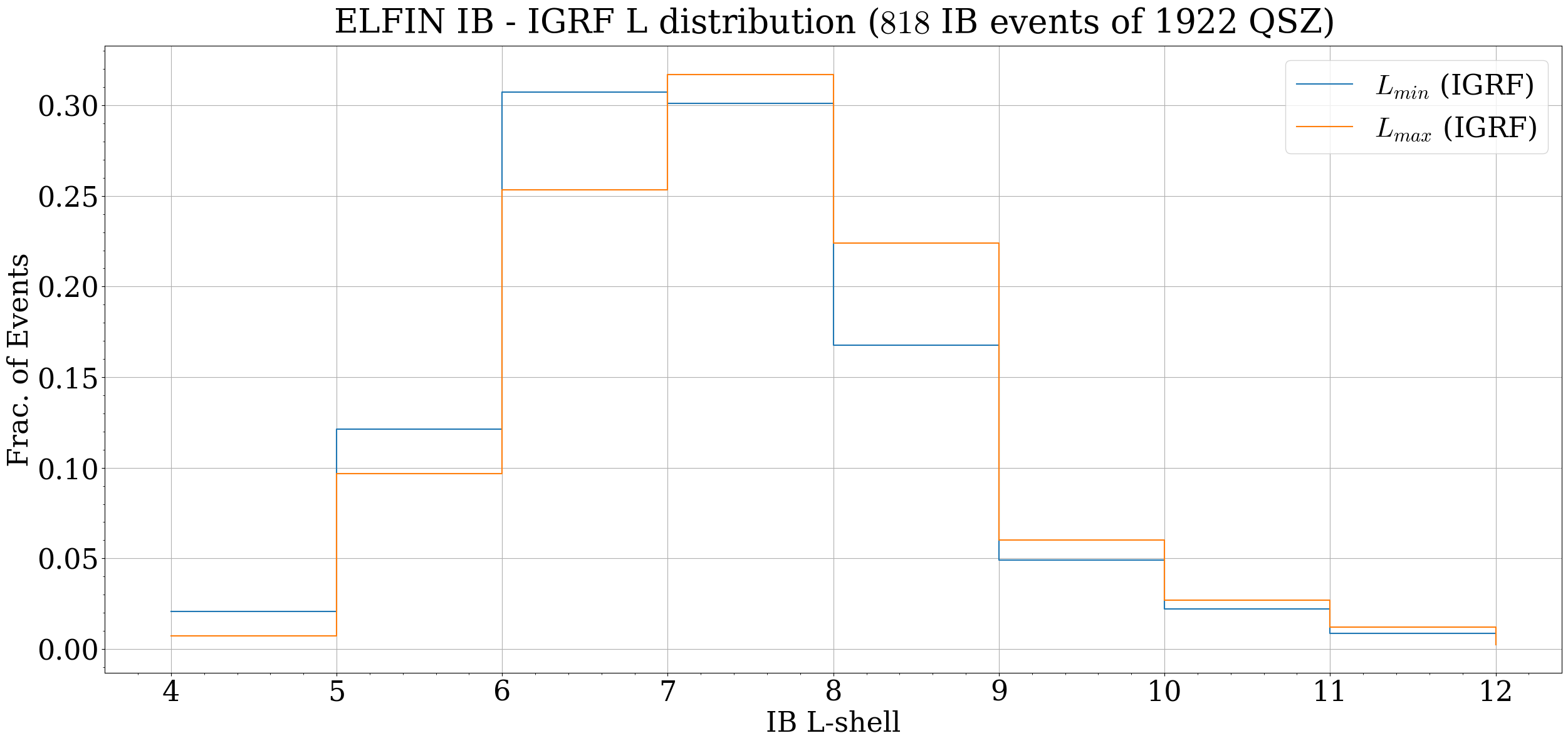}\\
\noindent\includegraphics[width=\textwidth]{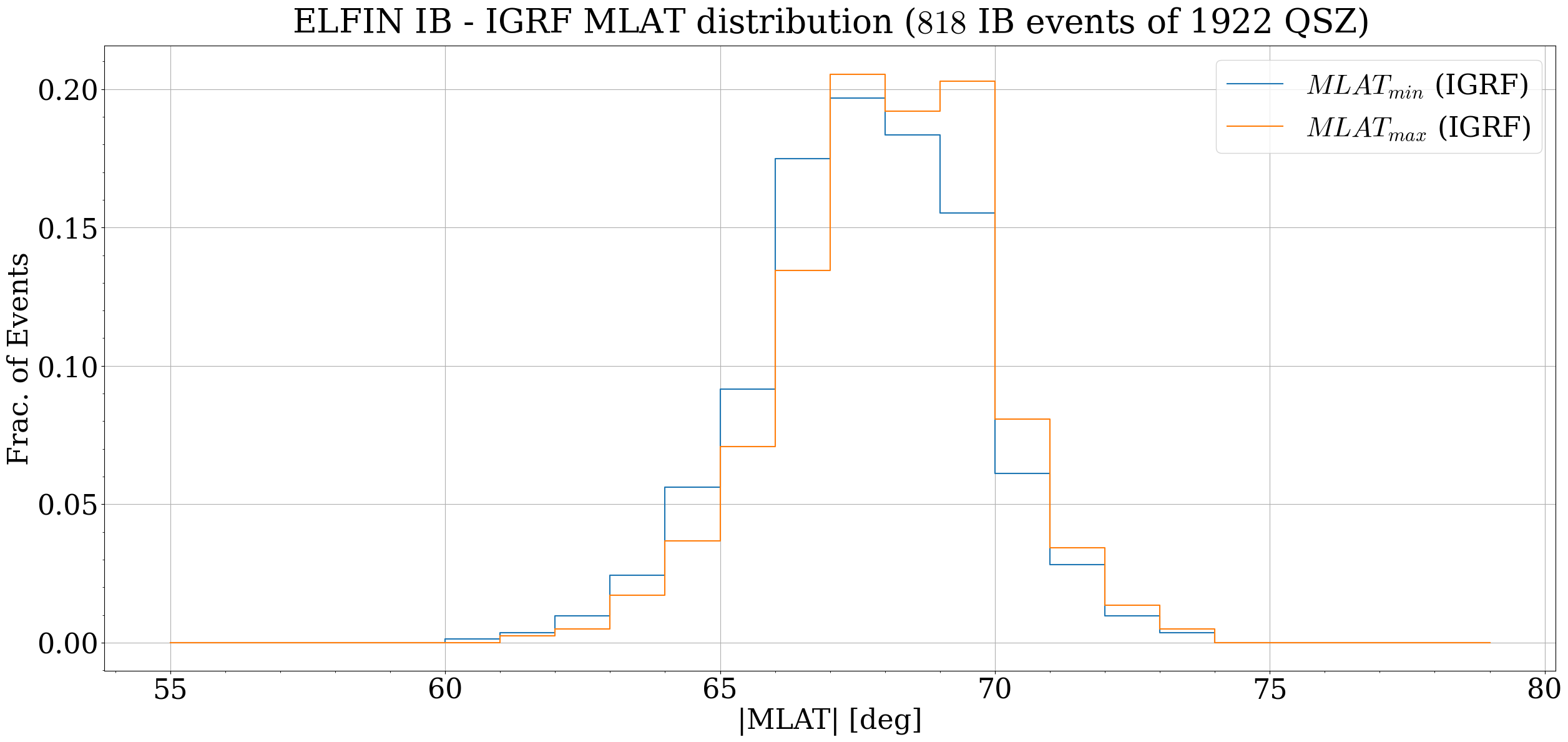}\\
\caption{Occurrence rates of IB crossings versus MLT (top), L-shell (middle), and magnetic latitude (bottom). Peak occurrence in MLT is at 22 MLT. Min and max refer to the low and high latitude boundary of the IB-dispersion signature. Median values for L-shell and MLAT are shown in the panels, and are L of 6-8 and 66$^\circ$-68$^{\circ}$, respectively, regardless of which boundary (low or high): the spread of these distributions due to geomagnetic activity and MLT is much larger than the thickness of the IB-dispersion. As a result the L-shell and MLAT distributions vary over wide ranges (several L-shells and $\sim$7$^\circ$ in MLAT around the respective medians).}
\label{Figure 4. }
\end{figure}

\begin{figure}
\noindent\includegraphics[width=\textwidth]{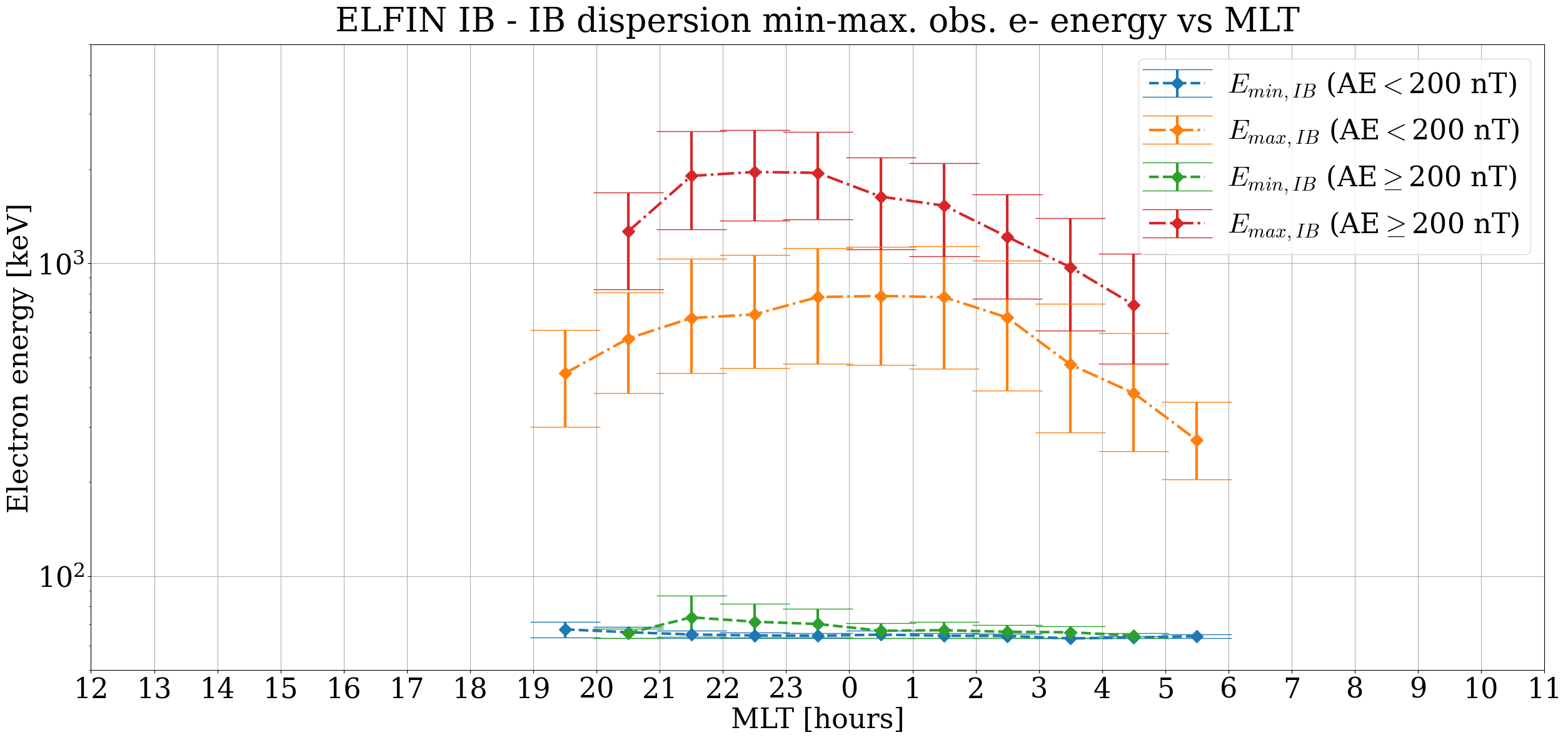}\\
\hspace{-1cm}\noindent\includegraphics[width=0.6\textwidth]{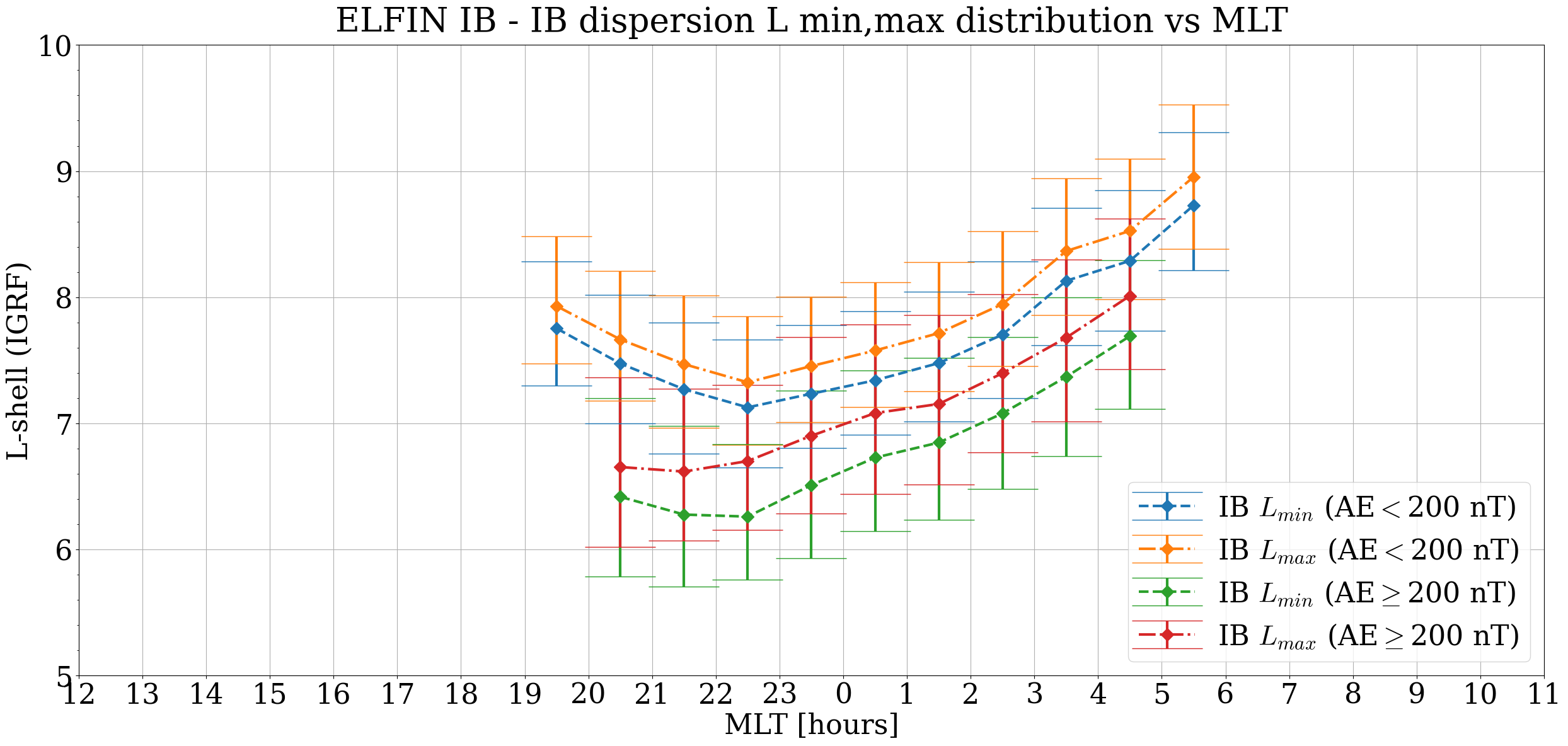}
\noindent\includegraphics[width=0.6\textwidth]{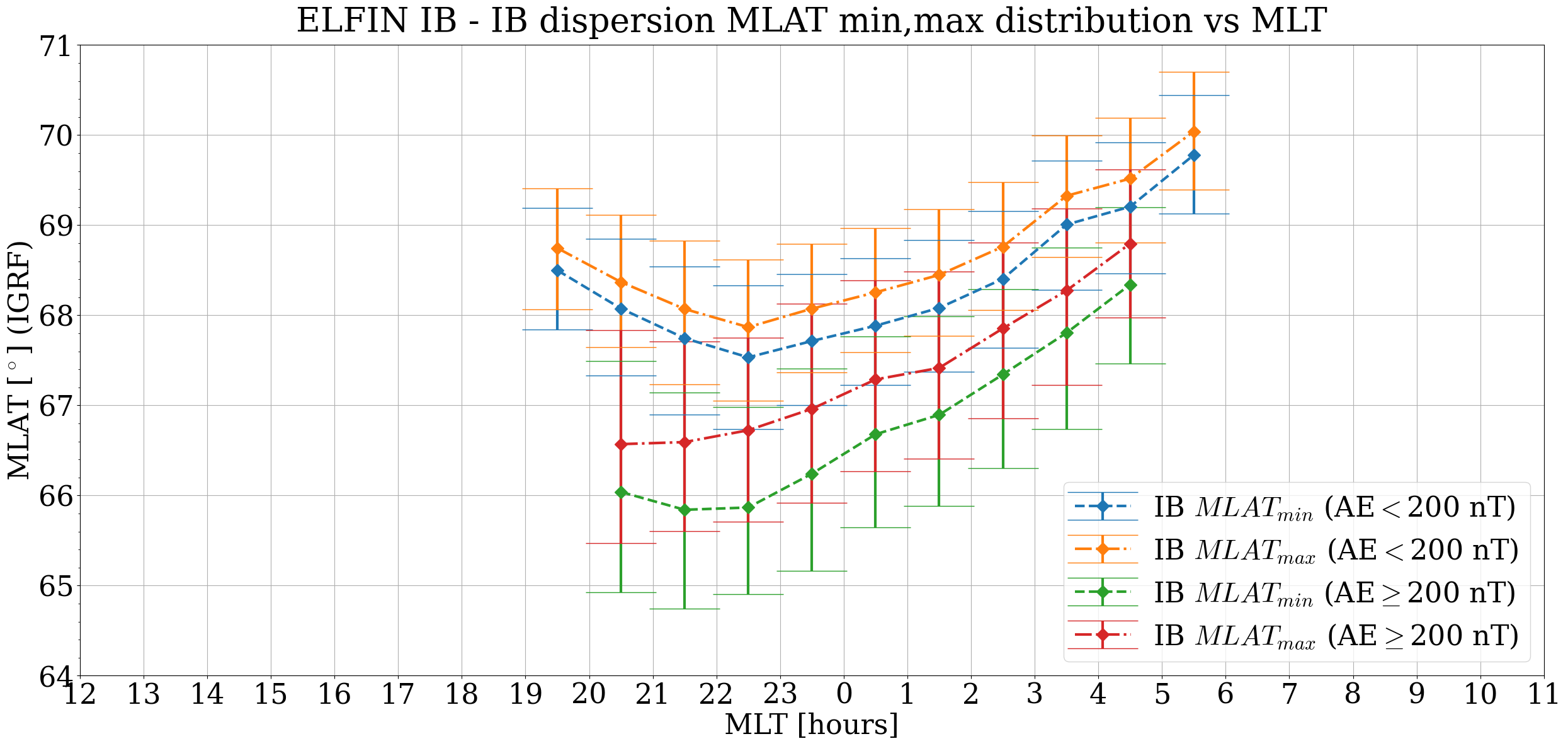}\\
\hspace{-1cm}\noindent\includegraphics[width=0.6\textwidth]{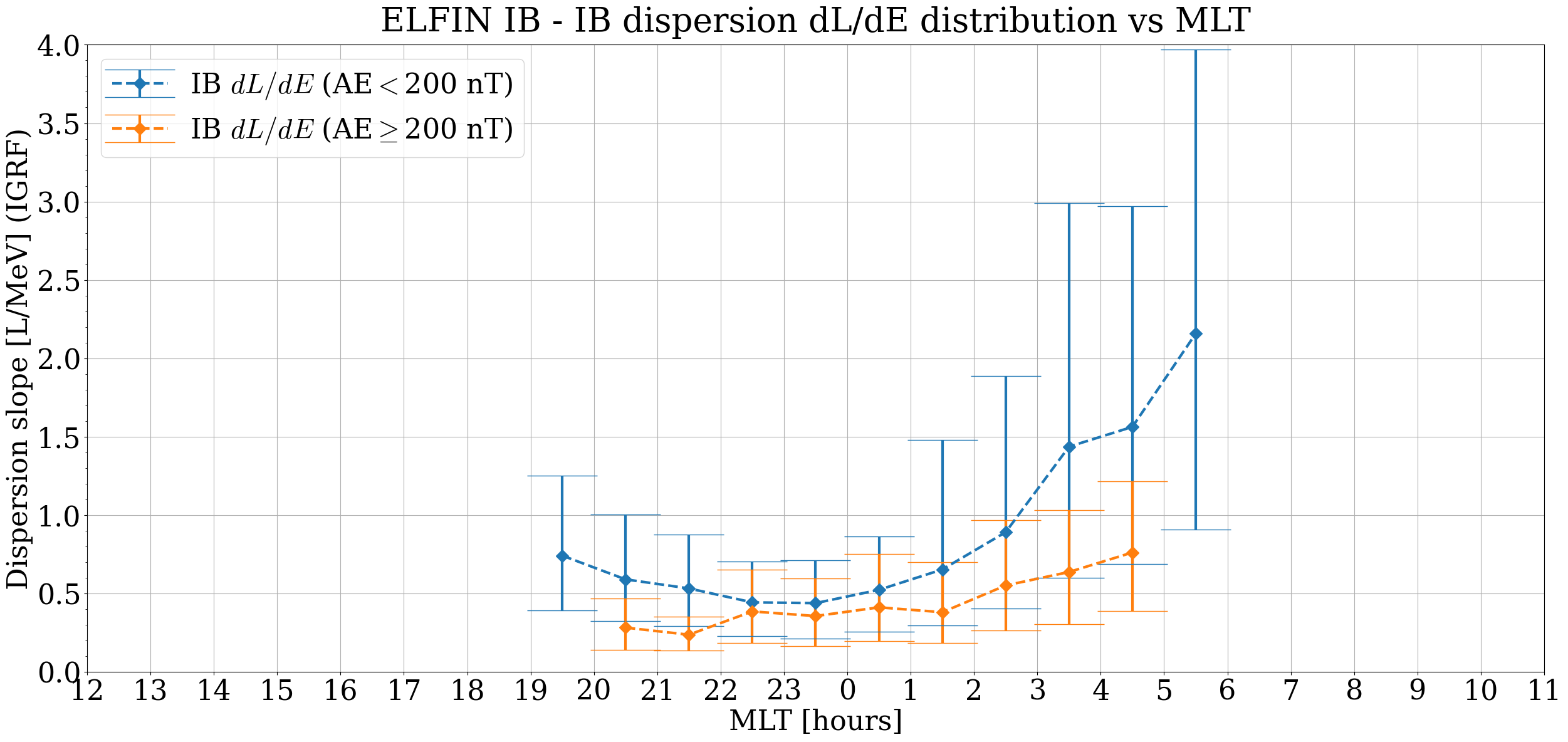}
\noindent\includegraphics[width=0.6\textwidth]{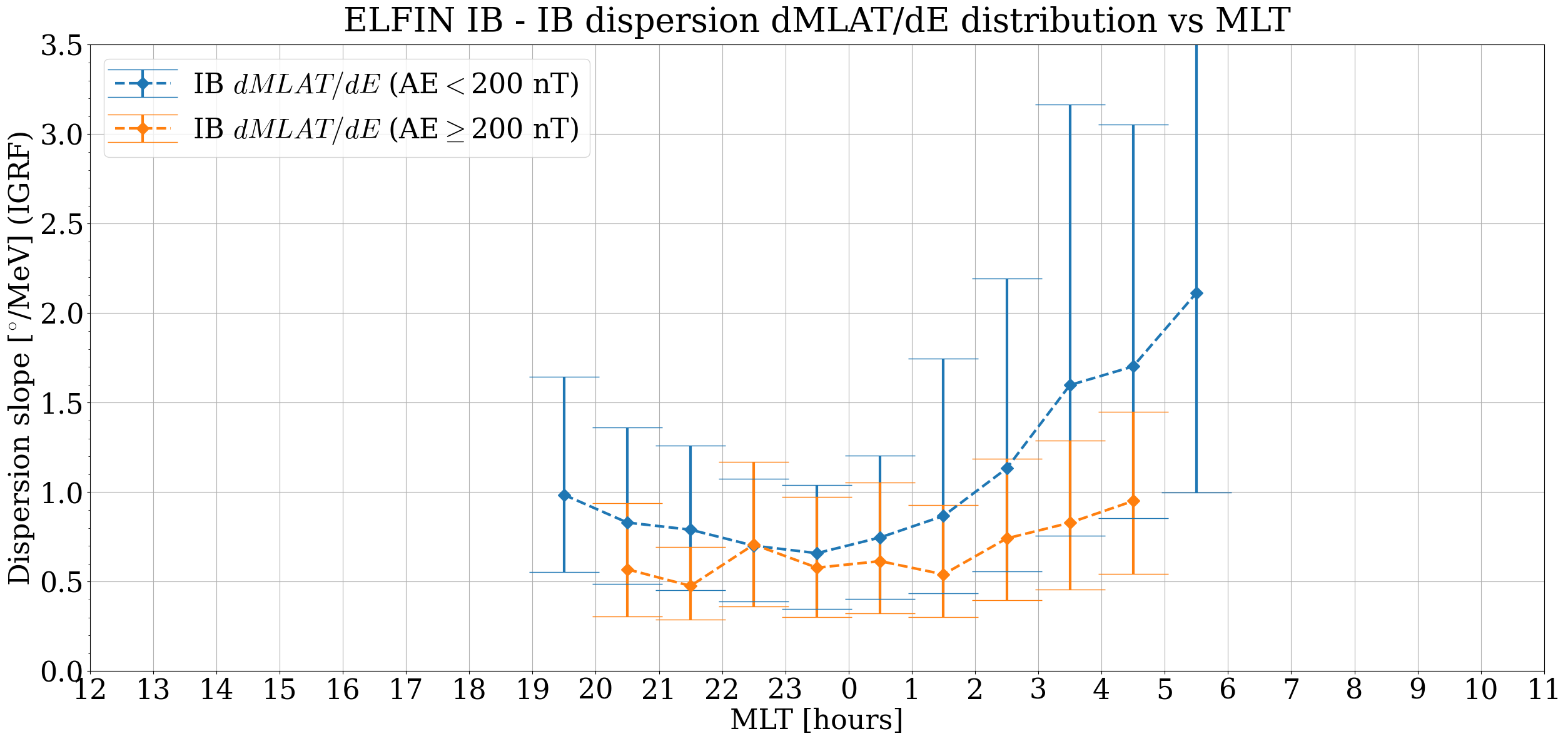}
\caption{Activity-dependent mean and probability-weighted standard deviation of IB dispersion region properties, including minimum and maximum energies appearing in the dispersion (top), minimum and maximum L-shells and magnetic latitudes (center), and slope of the dispersion in terms of L-shell and magnetic latitude (bottom). The general trend with increasing geomagnetic activity is for IBs to appear at lower latitudes and shift toward pre-midnight, accompanied by an increase in maximum electron energies. MLTs for which fewer than 5 events were observed were left blank.}
\label{Figure 5. }
\end{figure}

\begin{figure}
\noindent\includegraphics[width=\textwidth]{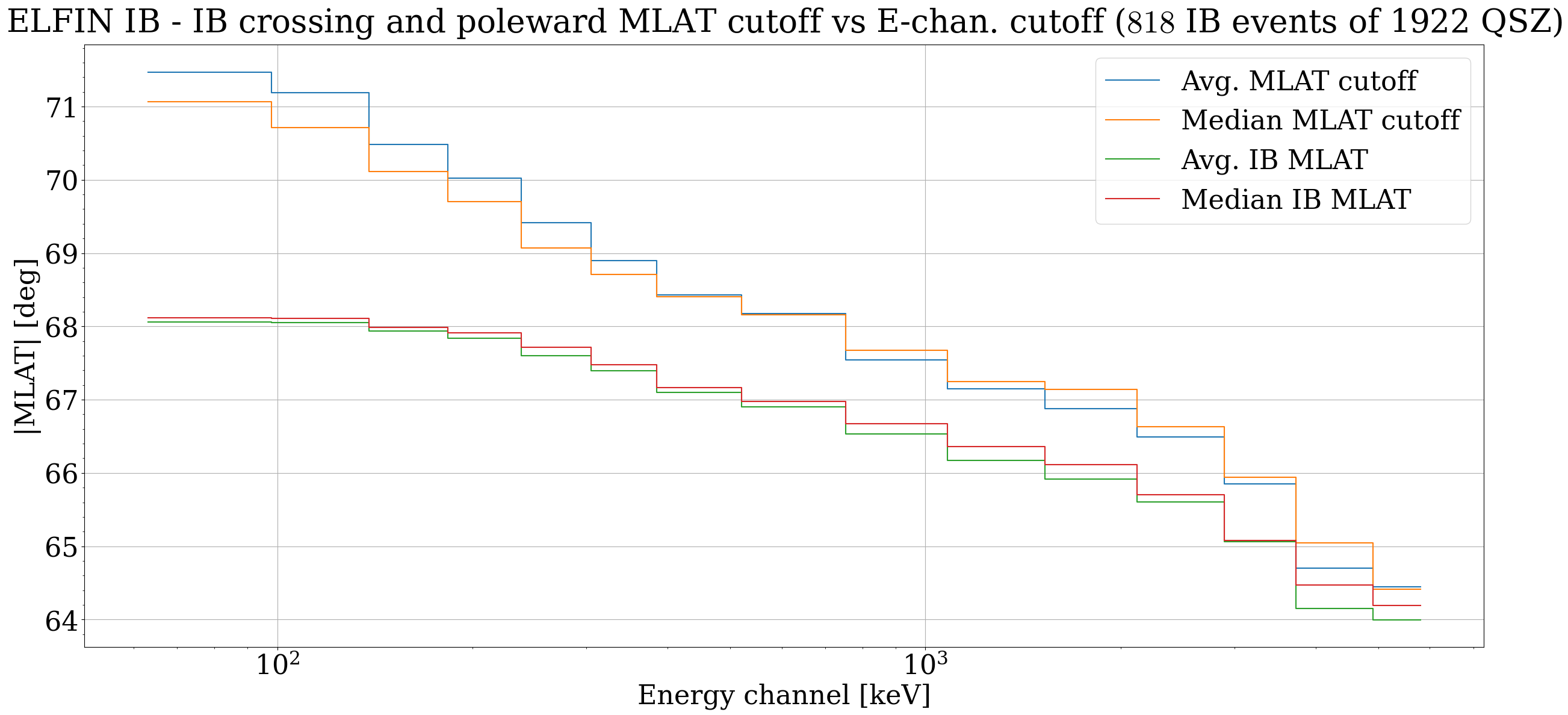}\\
\noindent\includegraphics[width=\textwidth]{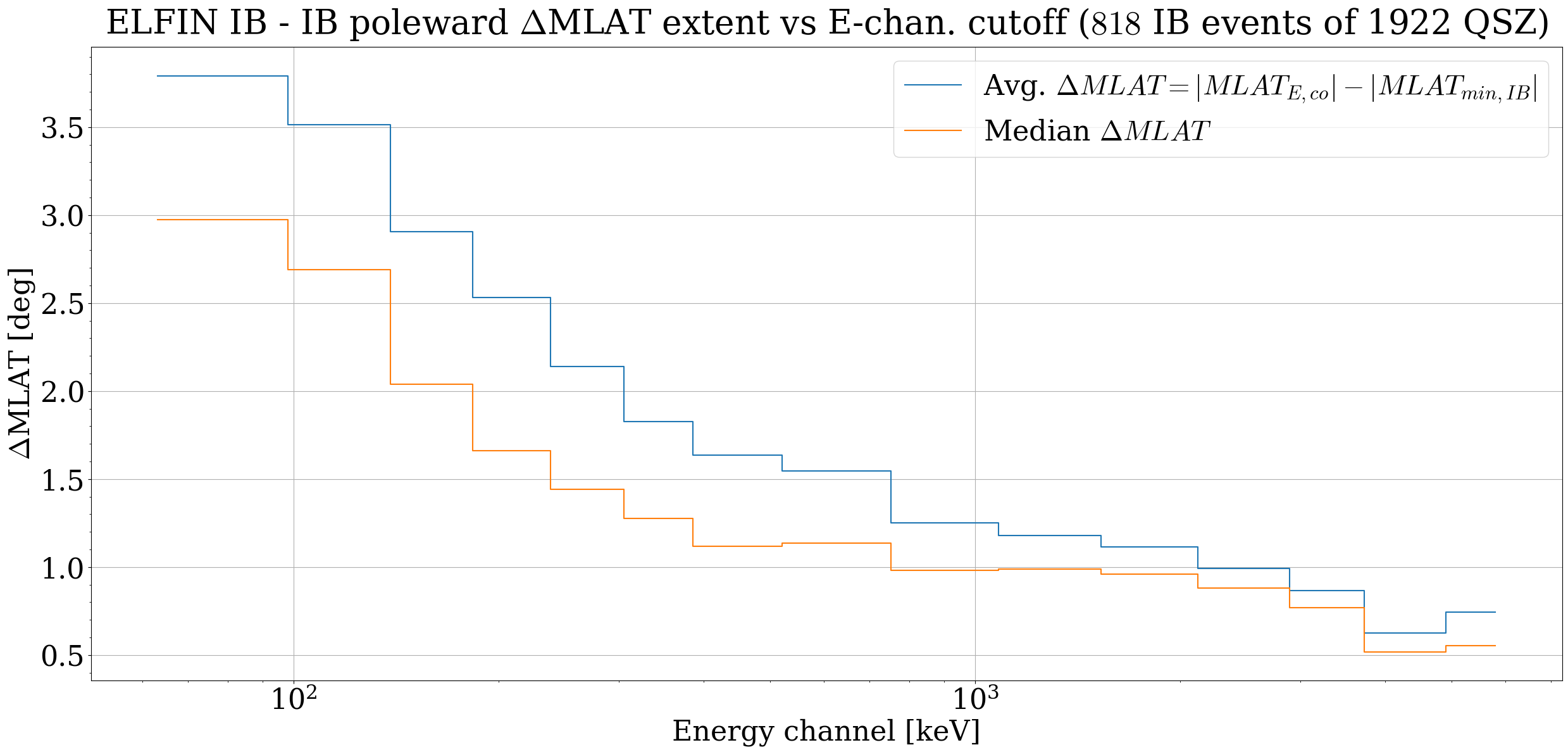}\\
\noindent\includegraphics[width=\textwidth]{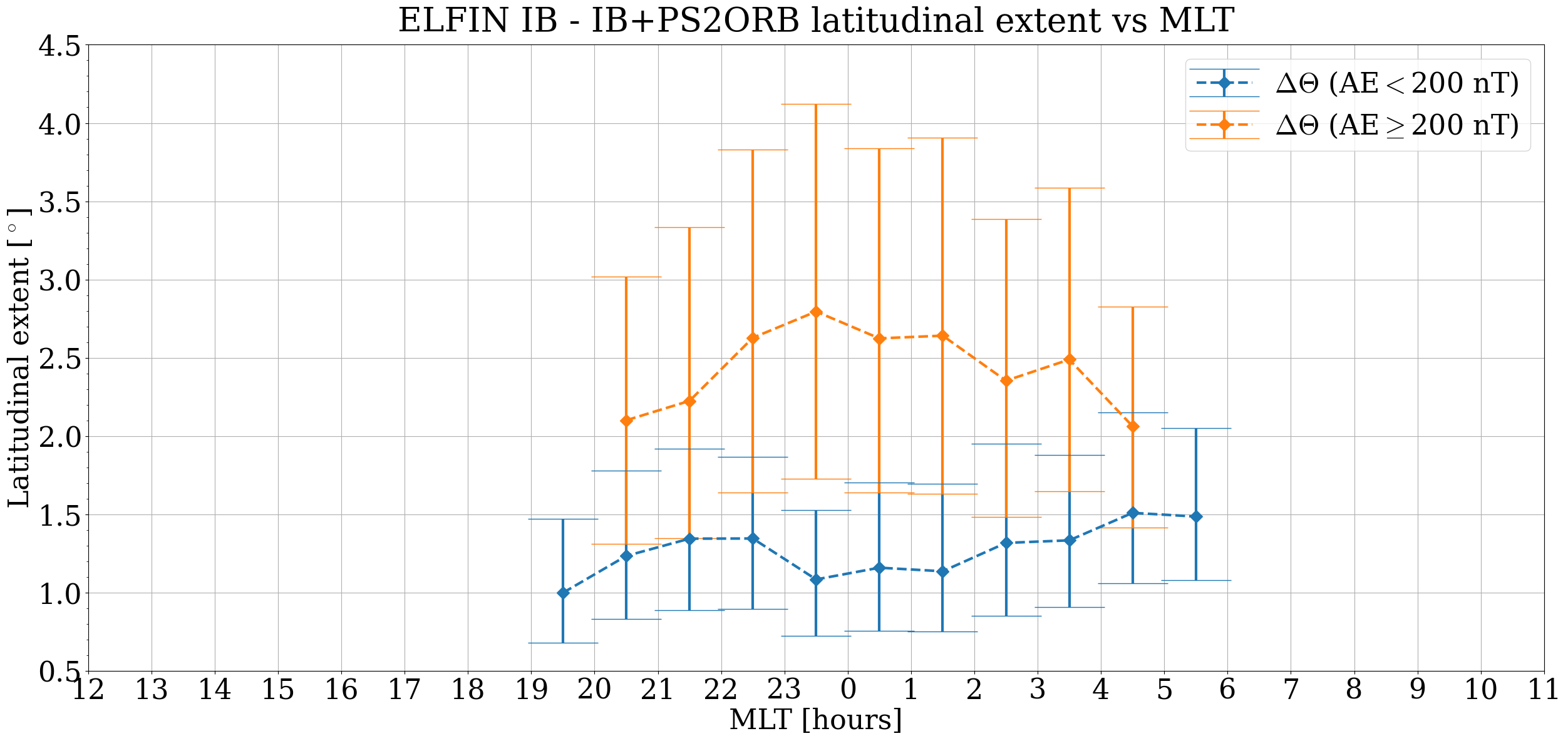}
\caption{Average absolute ELFIN-observed latitudinal profile of IB crossing locations and omnidirectional flux cutoff versus electron energy alongside average latitudinal extent from the IB crossing to omnidirectional flux cutoff versus energy (top), alongside the event-wise difference between these quantities (middle). A clear break in slopes is observed in the vicinity of 300 keV, suggesting a transition from the region of electrons dominated by field-line curvature scattering to that of the electron plasma sheet population (operationally used to define the latitudinal extent of the ``PS2ORB'' region). The mean latitudinal extent of the PS2ORB region in each MLT bin using the 300 keV cutoff criterion is shown in the bottom panel. The latitudinal width of the region dominated by FLCS exhibits clear variations with local time and activity.}
\label{Figure 6. }
\end{figure}

\begin{figure}
\noindent\includegraphics[width=\textwidth]{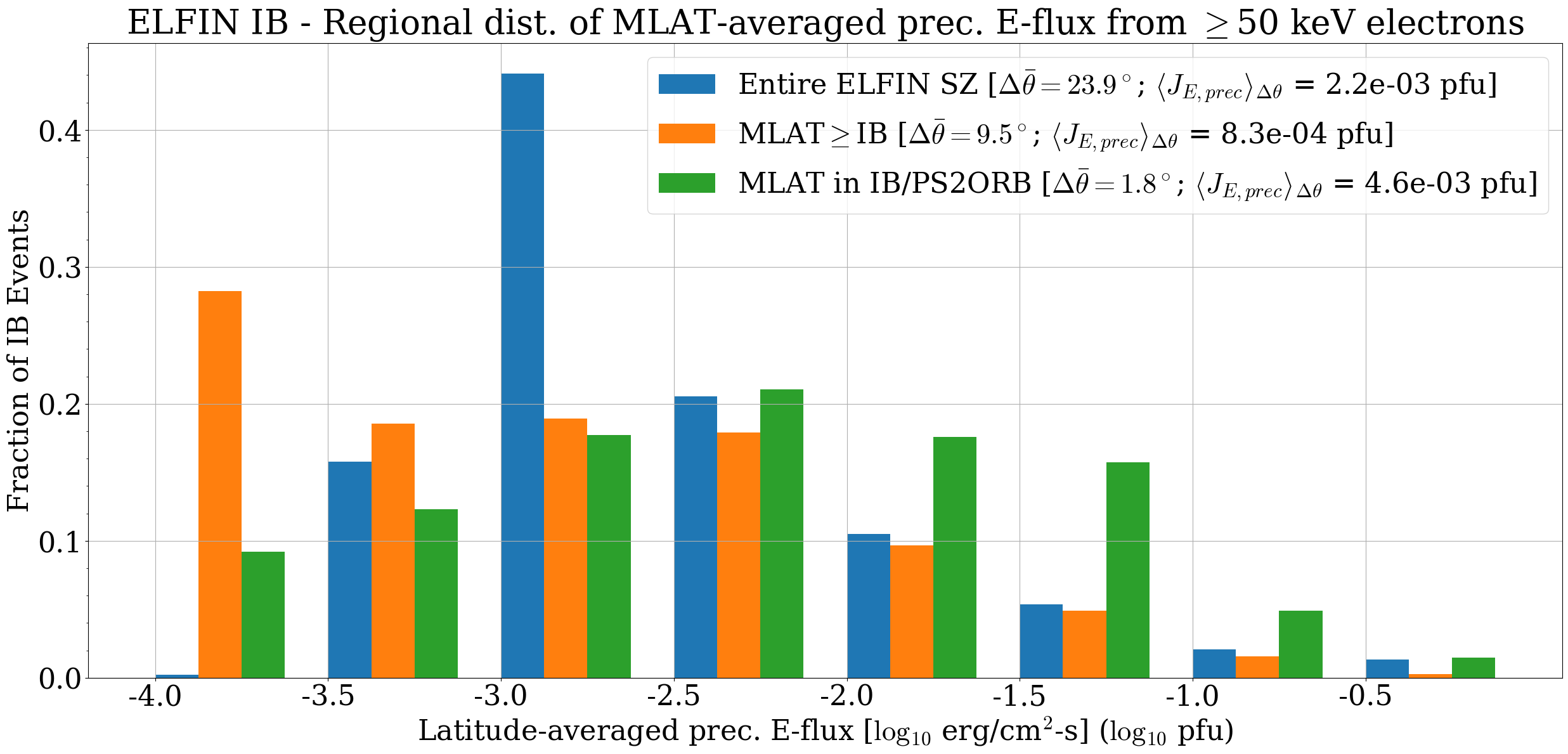}
\caption{Distribution of latitude-averaged precipitating energy flux from $\geq$50 keV electrons for three categories: The entire ELFIN science zone (blue); IB crossing and latitudes poleward of it (orange); and PS2ORB interface region (encompasses IB crossing). The quantity $\Delta \bar{\theta}$ represents the average latitudinal extent of the ELFIN observations in each region, while the corresponding quantity $\left<J_{E,prec} \right>_{\Delta \theta}$ represents the mean latitude-averaged integral energy flux over all events in erg/cm$^2$-s (or ``pfu''). The plot reveals that the IB-associated latitudes have the highest average energy flux of the latitude-mapped source regions under consideration.}
\label{Figure 7. }
\end{figure}

\begin{figure}
\noindent\includegraphics[width=\textwidth]{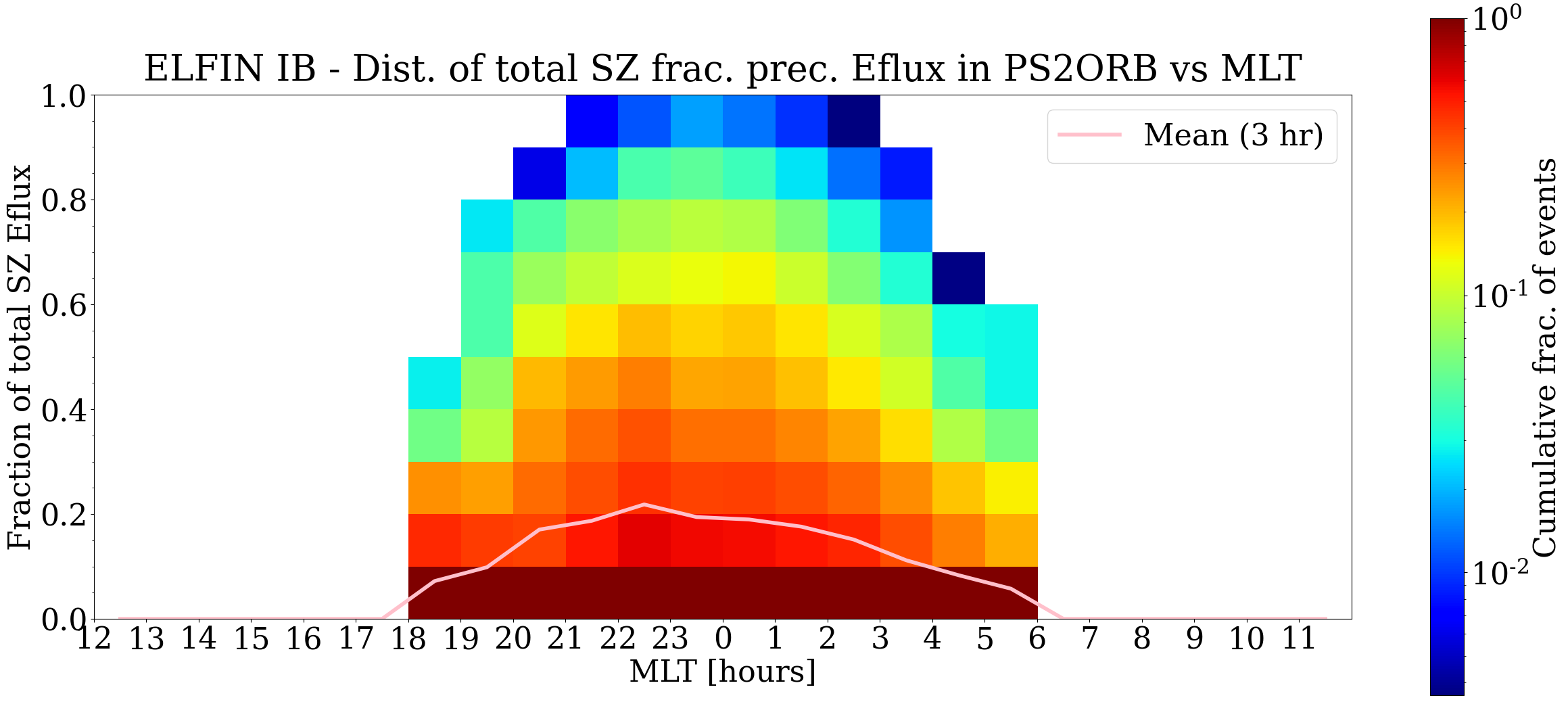}\\
\noindent\includegraphics[width=\textwidth]{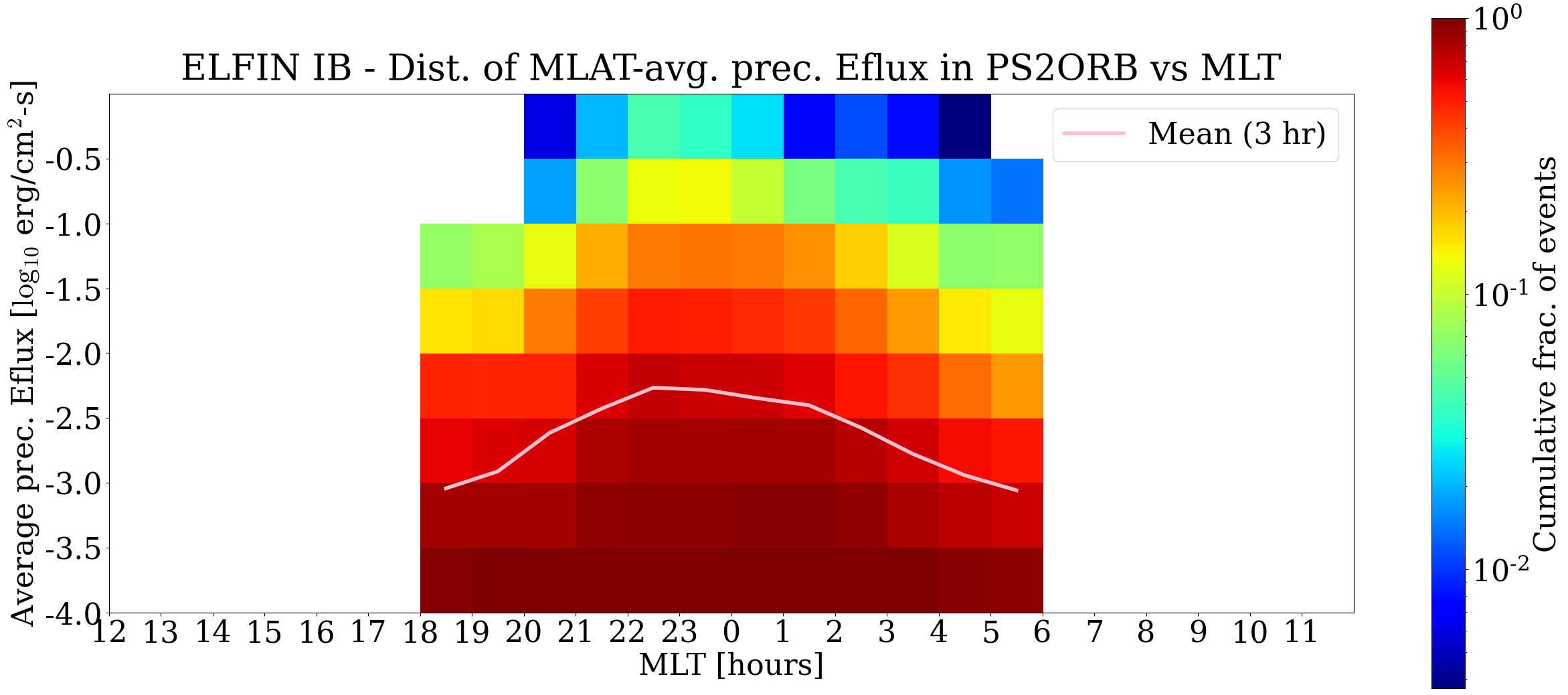}
\caption{Distribution of the precipitating kinetic energy flux of $\geq$50 keV electrons in the PS2ORB region versus MLT. Top: Fraction of the average total precipitating electron energy flux $\geq$50 keV compared with all latitudes observed in ELFIN science zones possessing an IB crossing. Bottom: average total precipitating energy flux from precipitating electrons versus MLT. The values were aggregated over all activity levels in the dataset, with each column representing a separate MLT-binned cumulative row-wise probability of occurrence.}
\label{Figure 8. }
\end{figure}

\begin{figure}
\noindent\includegraphics[width=0.33\textwidth]{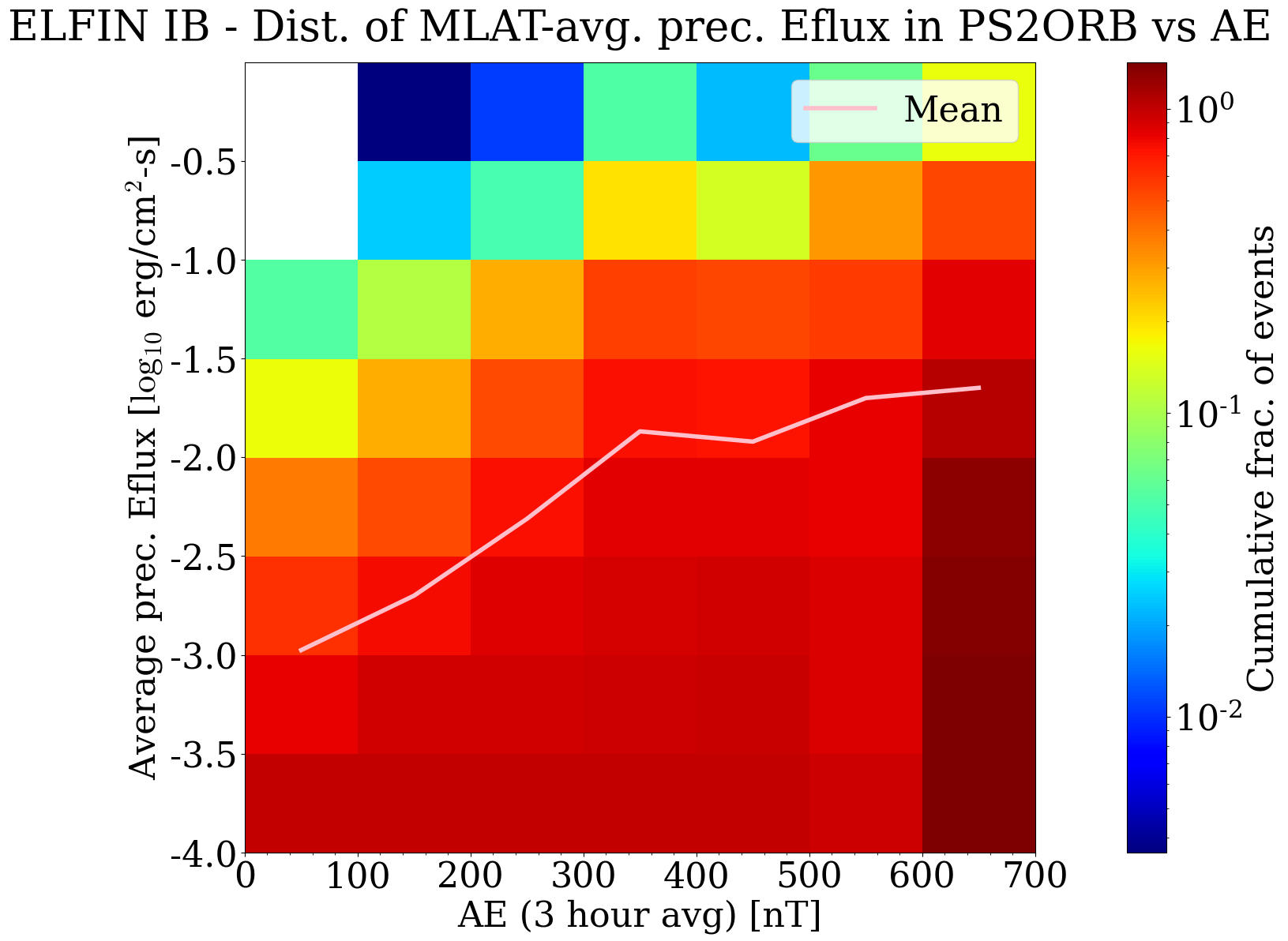}
\noindent\includegraphics[width=0.33\textwidth]{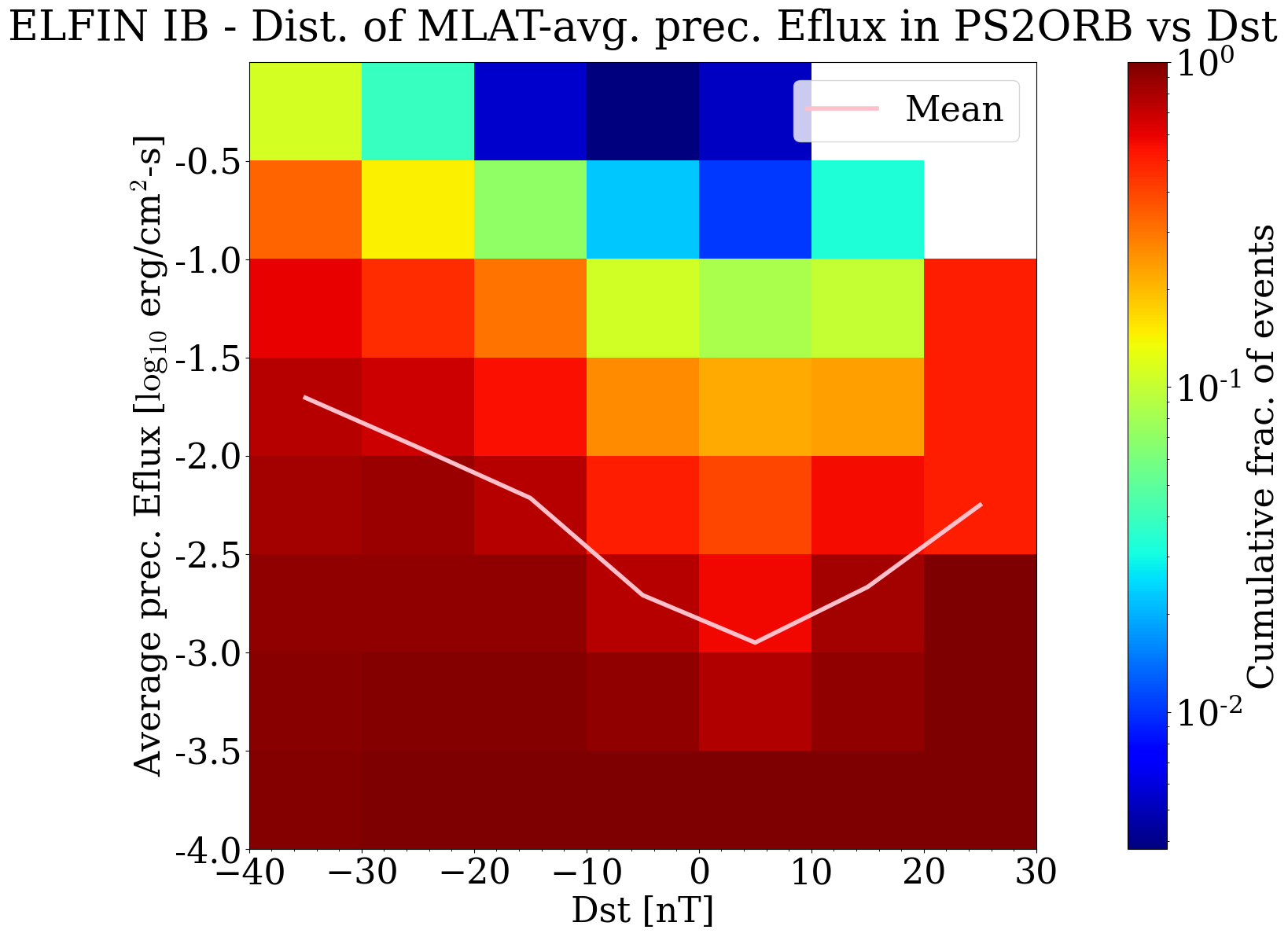}
\noindent\includegraphics[width=0.33\textwidth]{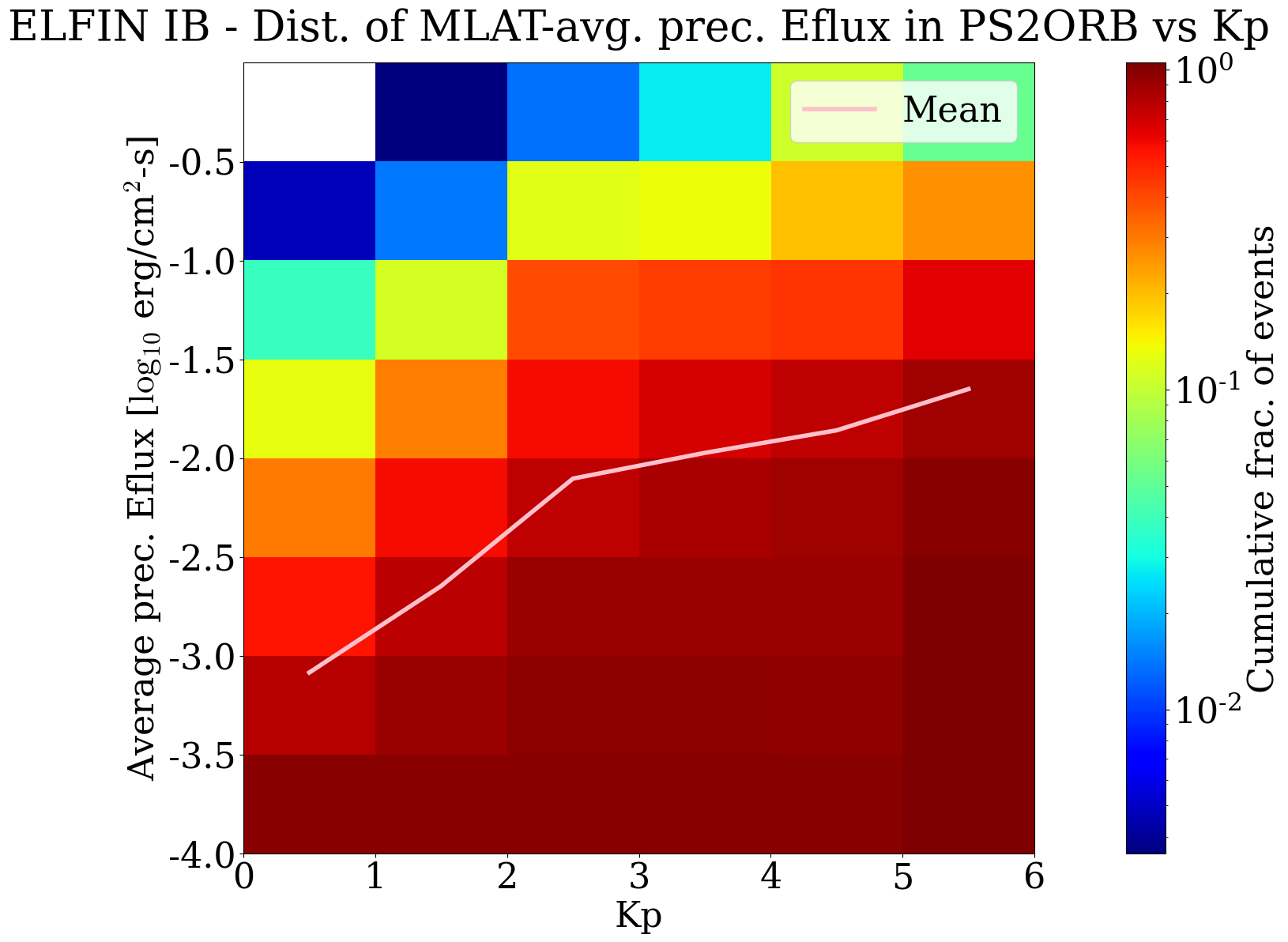}\\
\noindent\includegraphics[width=0.33\textwidth]{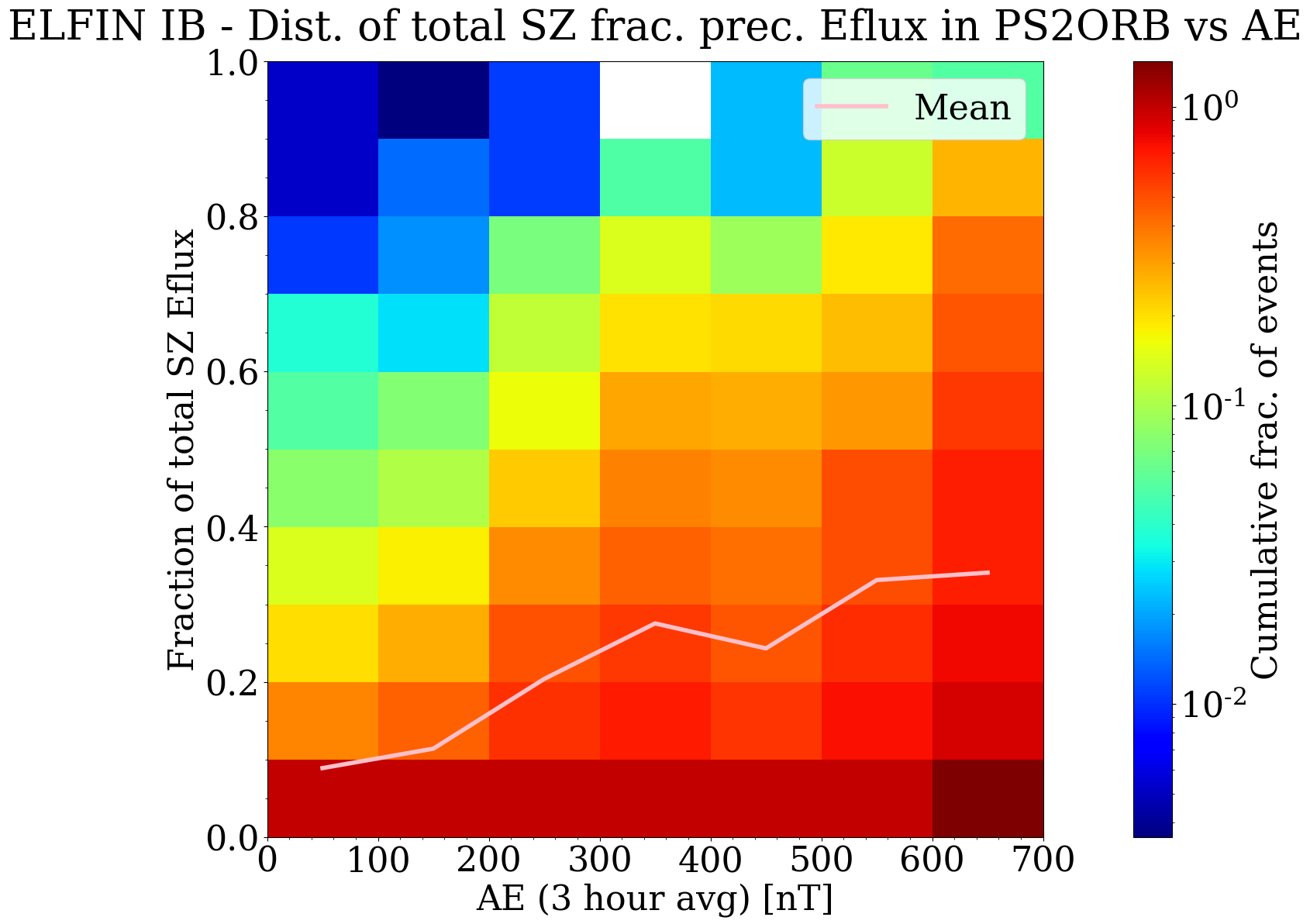}
\noindent\includegraphics[width=0.33\textwidth]{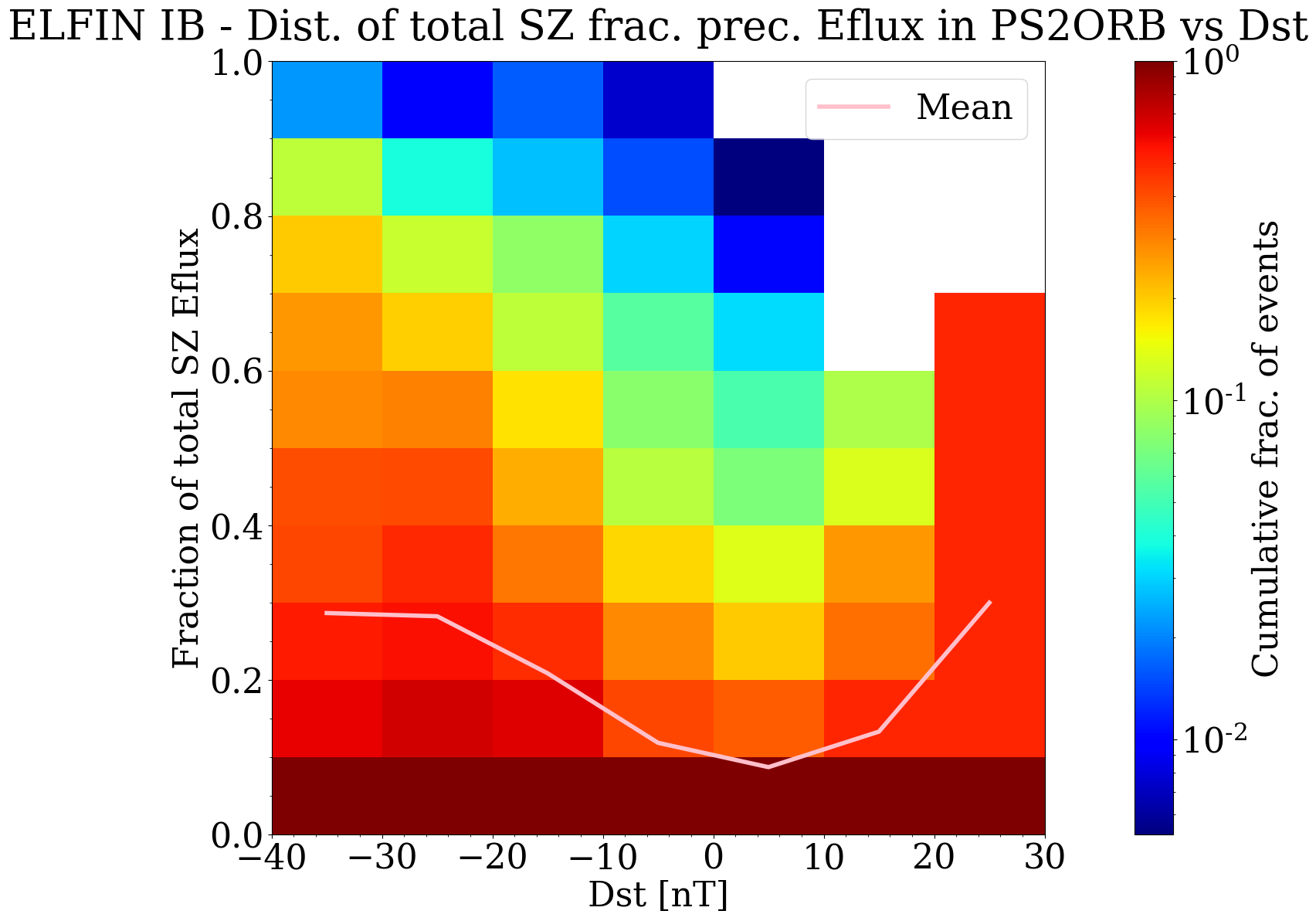}
\noindent\includegraphics[width=0.33\textwidth]{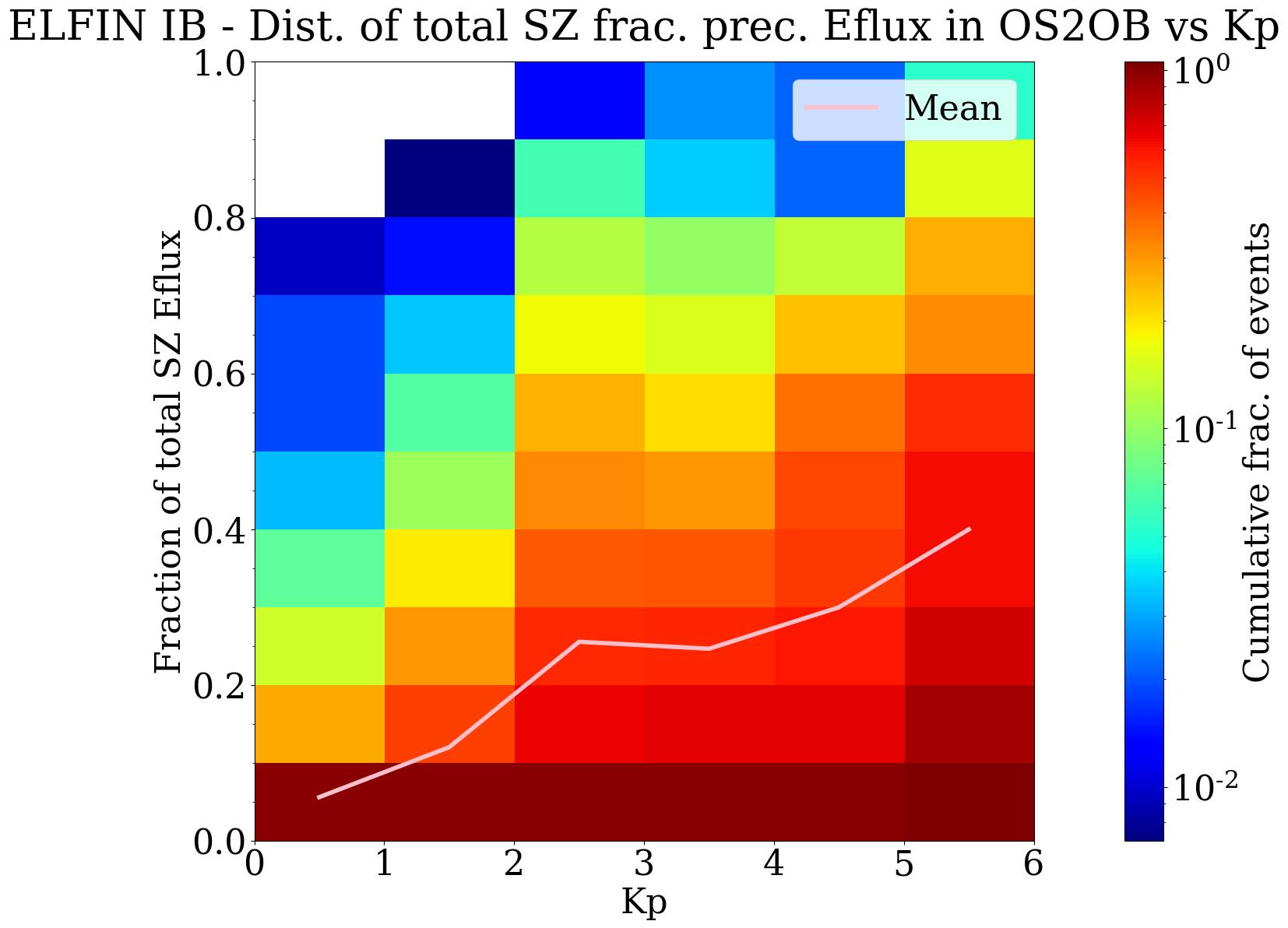}
\caption{Extension of Figure 7, binning by the geomagnetic activity indices AE (3 hour mean; left), Dst (middle), and Kp (right), aggregated over all MLTs containing IBs. Both the total and relative intensities of precipitation contributed by IBs/FLCS are observed to increase with activity, suggesting they can become even more significant during geomagetically-disturbed intervals.}
\label{Figure 9. }
\end{figure}


\end{document}